\documentclass{aa}

\usepackage[varg]{txfonts}
\usepackage{natbib}
\usepackage{graphicx}
\usepackage{dcolumn,url,longtable}
\usepackage{amssymb}
\usepackage{color}
\usepackage{tikz}
\usetikzlibrary{shapes,arrows}
\newcolumntype{d}{D{.}{.}{-1}}

\usepackage{float}
\usepackage{placeins}
\usepackage{capt-of}

\newcommand{\iu}{{i\mkern1mu}}

\begin{document}

\title{An advanced multipole model of the (130) Elektra quadruple system}
\titlerunning{An advanced multipole model of the (130) Elektra quadruple system}

\author{
M.~Fuksa\inst{\ref{auuk}} \and
M.~Brož\inst{\ref{auuk}} \and
J.~Hanuš\inst{\ref{auuk}} \and
M.~Ferrais\inst{\ref{lam}} \and
P.~Fatka\inst{\ref{auav}} \and
P.~Vernazza\inst{\ref{lam}} 
}

\institute{
    Charles University, Faculty of Mathematics and Physics, Astronomical Institute, V Hole\v{s}ovi\v{c}k\'{a}ch 2, 18000 Prague, Czech Republic\label{auuk}\\
    \email{fuksa@sirrah.troja.mff.cuni.cz}
      \and 
     Aix Marseille Univ, CNRS, LAM, Laboratoire d'Astrophysique de Marseille, Marseille, France\label{lam}
      \and 
    Astronomical Institute of the Czech Academy of Sciences, Fri\v{c}ova 298, Ond\v{r}ejov CZ-25165, Czech Republic\label{auav}
    }

\date{Received ??? / Accepted ???}

\abstract
{
The Ch-type asteroid (130)~Elektra is orbited by three moons, making it the first quadruple system in the main asteroid belt.
} 
{
We aim to characterise the irregular shape of Elektra and construct a complete orbital model of its unique moon system.
} 
{
We applied the All-Data Asteroid Modelling (ADAM) algorithm to
60~light curves of Elektra, including our new measurements,
46~adaptive-optics (AO) images obtained by the VLT/SPHERE and Keck/Nirc2 instruments, and 
two stellar occultation profiles. 
For the orbital model, we used an advanced $N$-body integrator,
which includes a multipole expansion of the central body (with terms up to the order $\ell = 6$),
mutual perturbations,
internal tides,
as well as the external tide of the Sun acting on the orbits.
We fitted the astrometry measured with respect to the central body
and also relatively, with respect to the moons themselves.
} 
{
We obtained a revised shape model of Elektra with the volume-equivalent diameter $(201\pm 2)\,{\rm km}$.
Out of two pole solutions,
$(\lambda, \beta) = (189; -88)\,{\rm deg}$ is preferred,
because the other one leads to an incorrect orbital evolution of the moons.
We also identified the true orbital period of the third moon S/2014~(130)~2 as
$P_2 = (1.642112 \pm 0.000400)\,{\rm d}$,
which is in between the other periods,
$P_1 \simeq 1.212\,{\rm d}$,
$P_3 \simeq 5.300\,{\rm d}$,
of S/2014~(130)~1 and S/2003~(130)~1, respectively.
The resulting mass of Elektra,
$(6.606 \substack{+0.007 \\ -0.013}) \times 10^{18}\,{\rm kg}$,
is precisely constrained by all three orbits.
Its bulk density is then
$(1.536 \pm 0.038)\,{\rm g\,cm}^{-3}$.
The expansion with the assumption of homogeneous interior leads to the oblateness
$J_2 = -C_{20} \simeq 0.16$.
However, the best-fit precession rates indicate a slightly higher value,
${\simeq}\,0.18$.
The number of nodal precession cycles over the observation time span 2014--2019
is 14, 7, and 0.5, respectively.
} 
{
Future astrometric or interferometric observations of Elektra's moons
should constrain these precession rates even more precisely
to track down possible inhomogeneities in primitive asteroids.
}


\maketitle


\section{Introduction}

The asteroid (130)~Elektra belongs to the primitive Ch spectral type objects \citep{Rivkin_2015AJ....150..198R}. The visible and near-infrared spectral data of Ch- and Cgh-type asteroids are most similar to the unheated CM chondrite meteorites, thus most likely representing their parent bodies \citep{Vilas1989, Vernazza2016}. The only minor spectral variations among the largest Ch/Cgh bodies and the members of the associated collisional families are likely due to differences in the average grain size of the regolith particles only. Therefore, CM parent bodies had homogeneous internal structures and did not experience significant ($>$300$^\circ$C) heating due to thermal evolution \citep{Vernazza2016}.
Yet, this body is not intact, because it experienced a collision,
which is commonly the origin of small satellites \citep{Durda_2004Icar..170..243D,Benavidez_2012Icar..219...57B}.

Elektra is orbited by three satellites (or, in other words, moons):
S/2003~(130)~1 \citep{marchis2008main},
S/2014~(130)~1 \citep{yang2016extreme},
S/2014~(130)~2 \citep{Berdeu_2022A&A...658L...4B}.
The diameters of the moons as derived from their photometry
are of the order of 6, 2, and 1.6\,km.
For the third moon,
a short-periodic orbit was found,
with a period of only 0.67\,d \citep{Berdeu_2022A&A...658L...4B}.

Previous shape modelling of Elektra \citep{hanuvs2017shape,vernazza2021vlt}
indicates that the ecliptic latitude $\beta$ is close to $-88^\circ$ or $-89^\circ$.
Hence, the ecliptic longitude $\lambda$ remains poorly constrained,
because even distant values of $\lambda$ represent similar directions in space.

In this work,
we revise the shape model
together with the orbits of the moons.
Most importantly,
this helps us to resolve the ambiguity of poles,
because the non-spherical shape of the central body
induces a precession of the pericentres $\omega$ and nodes $\Omega$,
which is constrained by the astrometry of the moons.
We also use additional photometric data, as described in Sect.~\ref{sec:data}

From now on,
bodies are denoted as follows:
0 .. (130) Elektra,
1 .. S/2014~(130)~1,
2 .. S/2014~(130)~2,
3 .. S/2003~(130)~1.
As explained in Sect.~\ref{sec:orbit}, this order corresponds
to the respective orbits:
1 .. inner,
2 .. middle, and
3 .. outer.
All physical and orbital quantities are thus numbered accordingly.


\section{Data}\label{sec:data}

\subsection{Lightcurves, AO images and stellar occultations of Elektra}

\begin{table}
\begin{center}
\caption{\label{tab:lcs}Summary of optical disk-integrated light curves of (130)~Elektra obtained in this work by the BlueEye600 robotic observatory. Each line represents one light curve and contains the epoch, the number of individual measurements $N_p$, the asteroid's distances to the Earth $\Delta$ and the Sun $r$, phase angle $\varphi$, photometric filter and observation information.}
\begin{tabular}{rlrrrrc}
\hline 
\multicolumn{1}{c} {N} & \multicolumn{1}{c} {Epoch} & \multicolumn{1}{c} {$N_p$} & \multicolumn{1}{c} {$\Delta$} & \multicolumn{1}{c} {$r$} & \multicolumn{1}{c} {$\varphi$} & \multicolumn{1}{c} {Filter}\\
 &  &  & (AU) & (AU) & (\degr) & \\
\hline\hline
    1  &  2022-03-01.2    &  112  &  2.66  &  3.62  &  4.3   &  R  \\
    2  &  2022-03-02.2    &  60   &  2.66  &  3.62  &  4.1   &  R  \\
    3  &  2022-03-03.2    &  39   &  2.65  &  3.62  &  3.9   &  R  \\
    4  &  2022-03-04.2    &  29   &  2.65  &  3.62  &  3.8   &  R  \\
    5  &  2022-03-10.2    &  71   &  2.66  &  3.63  &  3.5   &  R  \\
\hline
\end{tabular}
\end{center}
\end{table}

For the shape reconstruction of Elektra, we prepared a data set consisting of
60 light curves,
46 AO images, and
two stellar occultations.

Most of the light curves (55 out of 60) were taken from the Database of Asteroid Models from Inversion Techniques \citep[DAMIT\footnote{\url{https://astro.troja.mff.cuni.cz/projects/damit/}},][]{vdurech2010damit}
which besides shape models contains also light curves and other input parameters.
The observational data covered the time span from 1980--2016.
In March 2022, we expanded the set by measuring five new light curves (Table \ref{tab:lcs})
using the BlueEye600 robotic observatory \citep{vdurech2018shape}, located in Ondřejov.
These light curves were obtained exclusively for this study and are now available in DAMIT together with the new shape and spin state solution. Moreover, we also uploaded them to the ACLDEF\footnote{\url{https://alcdef.org/}} database.
All of the lightcurves used in this study, including the new measurements, can be found in Appendix \ref{app:lightcurves}.

The set of AO images consists of two subsets.
The first one from \citet{hanuvs2017shape},
consists of 14 images captured by the Keck/Nirc2 instrument during the years 2002--2012
and two high-resolution infrared images were taken by the AO system VLT/SPHERE in 2014.
The InfraRed Differential Imager and Spectrograph \citep[IRDIS,][]{dohlen2008infra} and
the Integral Field Spectrograph \citep[IFS,][]{claudi2008sphere}
were used simultaneously to cover a more extended spectral range.

The second sub-set of 30 images,
some of which can be seen in Fig.~\ref{fig:comparison},
is taken from \citet{vernazza2021vlt}.
It was obtained by the Zurich IMaging POLarimeter \citep[ZIMPOL,][]{schmid2018sphere}
during the summer of 2019.

\begin{figure}
  \resizebox{\hsize}{!}{
    \includegraphics[width=.45\linewidth]{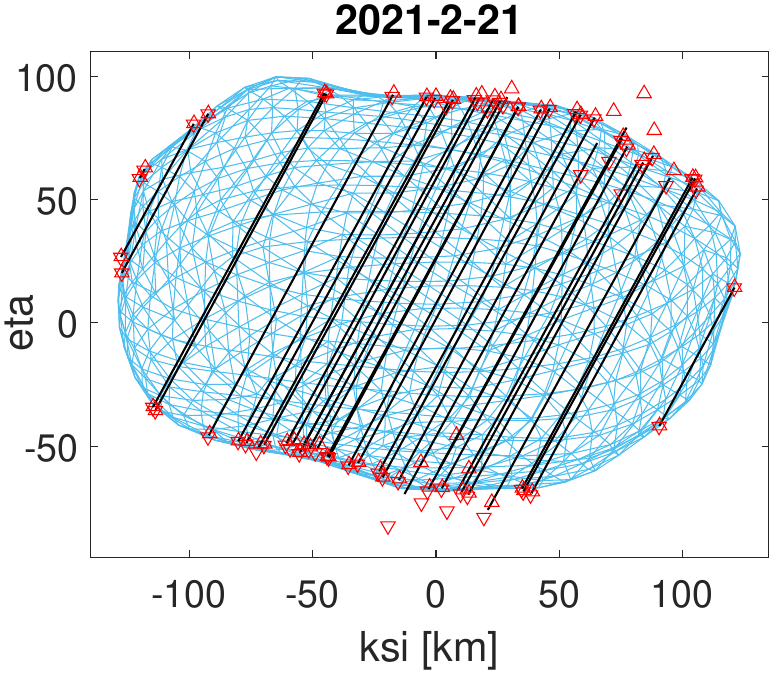}
    \hspace{.005\linewidth}
    \includegraphics[width=.45\linewidth]{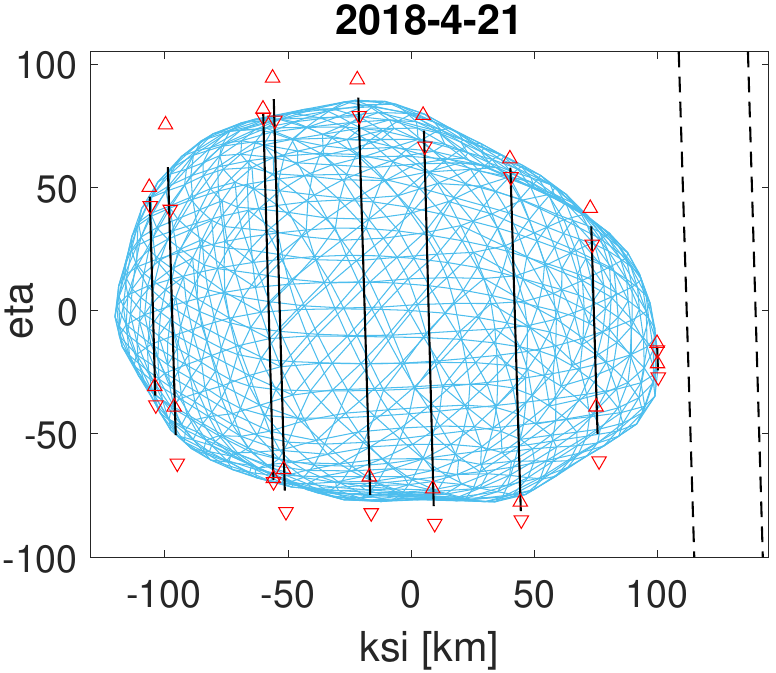}
    }
\caption{
Comparison of Elektra's shape model against chords from occultation events.
The \emph{red triangles} represent the timing uncertainties at the ends of each chord and
the \emph{dashed lines} are non-detection chords.}
\label{fig:occultations}
\end{figure}

The two stellar occultations were taken from
the Asteroidal Occultation Observers in Europe\footnote{\url{http://www.euraster.net/index.html}} database.
The first event occurred on the 21st of April 2018 and was observed by 48 observers in total.
After excluding 4 far misses and 6 chords with incorrect time indications, we were left with 38 usable chords.
The second event occurred on the 21st of February 2021 and was observed by 17 observers.
Eight of the chords were non-detections though.
Even though such chords are generally useful as `strict' bounds for the asteroid's shape,
we didn't use them because the shape is already well-constrained.
Both occultation events are shown in Fig.~\ref{fig:occultations}.

\subsection{Astrometric measurements of the moons}

For the process of fitting the orbital parameters of the moons,
it's crucial to have a sufficient set of positions,
usually related to the central body or its photocentre.
These positions were measured, with various methods,
on the AO images of Elektra,
whenever some of its moons were visible.

The first data set was taken from \citet{Berdeu_2022A&A...658L...4B},
where the process of reduction, halo removal and astrometry is explained in detail.
Their data set is based on individual AO images of Elektra from December 2014
and consists of
120 positions of S/2014~(130)~1,
120 positions of S/2014~(130)~2, and
150 positions of S/2003~(130)~1.
However, it is important to note that the individual positions are not mutually independent.
In fact, they correspond to a linear fit (in time) of 40 or 50 unique measurements.

The second data set is based on 30 AO images from \citet{vernazza2021vlt}.

The images were reduced using the eclipse data reduction package \citep{devillard1997eclipse}.
After reduction, the images were separately processed to remove the bright halo surrounding the asteroid, which was a necessary step to improve the detectability of the moons.
To obtain the astrometric positions of the satellites, we used a specialized algorithm \citep{2013Icar..226.1045H} that extracted the primary’s contour and determined its photocentre. By fitting a Moffat-Gauss source profile to the moons and measuring from the photocentre, we obtained the 
20, 20, and 12 new positions of the respective moons.
Moreover, we computed the astrometric positions of the moons
with respect to the other moons,
which is handy to avoid any systematic related to the photocentre of the central body,
or possibly centre-of-mass corrections.

The new positions are listed in Appendix \ref{app:observation_data}. Additionally, all astrometric data is available for download from the {\tt Xitau}%
\footnote{\url{https://sirrah.troja.mff.cuni.cz/~mira/xitau/}} webpage.


\section{Shape reconstruction}\label{sec:shape}

\subsection{All-Data Asteroid Modelling algorithm}

The ADAM algorithm \citep{viikinkoski2015adam, viikinkoski2016tampere} is a versatile inversion technique
for the shape reconstruction of asteroids from various data types (AO images, light curves, occultations, radar, etc.).
ADAM minimises the objective function,
which is a measure of the difference between the Fourier-transformed image and the projected shape:
\begin{equation}
\begin{split}
    \label{eq:ADAM_metric}
    \chi^2 & = w_{\rm AO}\! \sum_{i}^{} \sum_{j=1}^{N_i} \left\Vert \mathcal{F} D_i (u_{ij}, v_{ij}) - e^{2 \pi \iu (o_{i}^{x} u_{ij} + o_{i}^{y} v_{ij}) + s_i} \mathcal{F} M_i (u_{ij}, v_{ij}) \right\Vert^2\,+ \\
    & +\, w_{\rm LC}\kern1pt \chi_{\rm{LC}}^{2} +\, w_{\rm OC}\kern1pt \chi_{\rm{OC}}^{2} + \sum_{i}^{} w_i \gamma_{i}^{2}, 
\end{split}
\end{equation}
where $\mathcal{F} D_i (u_{ij}, v_{ij})$ is the Fourier transform of the data image~$D_i$ and
$\mathcal{F} M_i (u_{ij}, v_{ij})$ is the Fourier transform of the projected shape~$M_i$,
corresponding to the $i$-th image and evaluated at the $j$-th frequency point $(u_{ij}, v_{ij})$.
The scale $s_i$ and offset $(o_i^{x}, o_i^{y})$ are free parameters resolved during the optimisation.
The additional term $\chi_{\rm{LC}}^{2}$ is the standard square norm of the light curve fit, while the term $\chi_{\rm{OC}}^{2}$ is the model fit to stellar occultation chords. The implementation of the latter term in ADAM is described in \citet{hanuvs2017shape}.
The first three terms are multiplied by their corresponding weights $w_{\rm AO}$, $w_{\rm LC}$ and $w_{\rm OC}$ and   
the last term is a sum of regularisation functions~$\gamma_i$,
multiplied by the weights~$w_i$
\citep{viikinkoski2015adam,hanuvs2017shape}.

The data weights have an impact on the initial convergence of $\chi^2$ and subjectively depend on the dataset. They are chosen to balance the individual $\chi^2$ values and enable good convergence. Regularization weights then help guide the
convergence during the process.

Two shape representations are supported: octantoids and subdivisions.
Octantoids are a global parametrization,
whereas subdivisions offer more local control.
In this work, we used the octantoid shape representation.
For a more detailed explanation see \citet{viikinkoski2015adam}.

\subsection{Best-fit and alternative shape models}

We utilised a two-step approach.
First, we aimed to get just a coarse shape model.
Second, the shape resulting from the first optimisation was used as an initial model,
and we then doubled the number of facets and the spherical harmonics degree,
in order to capture surface details.
The initial conditions for the convergence,
which are needed for the period and the pole,
were taken from \citet{vernazza2021vlt}.

\begin{figure}
  \resizebox{\hsize}{!}{
    \includegraphics[width=.3\linewidth]{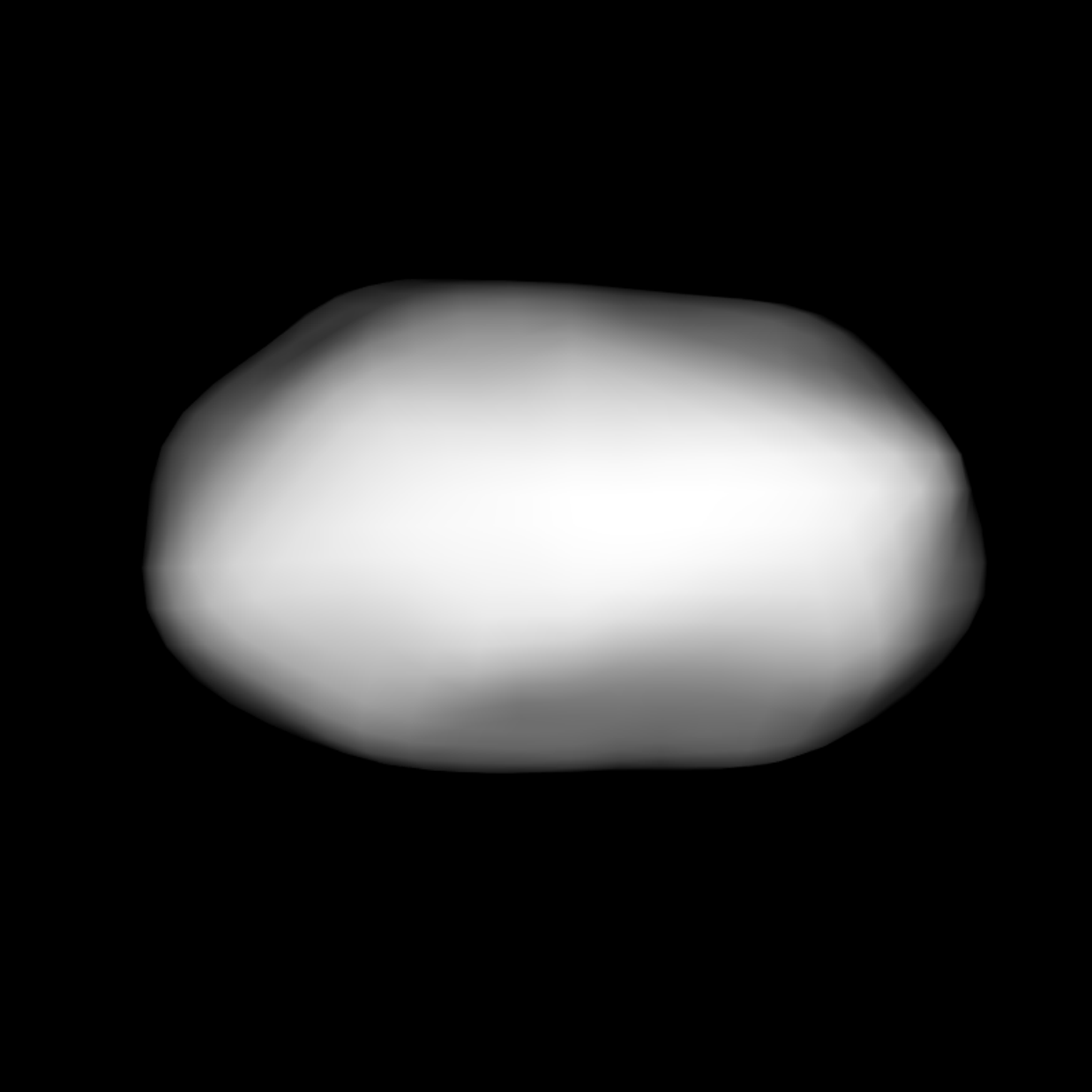}
    \hspace{.005\linewidth}
    \includegraphics[width=.3\linewidth]{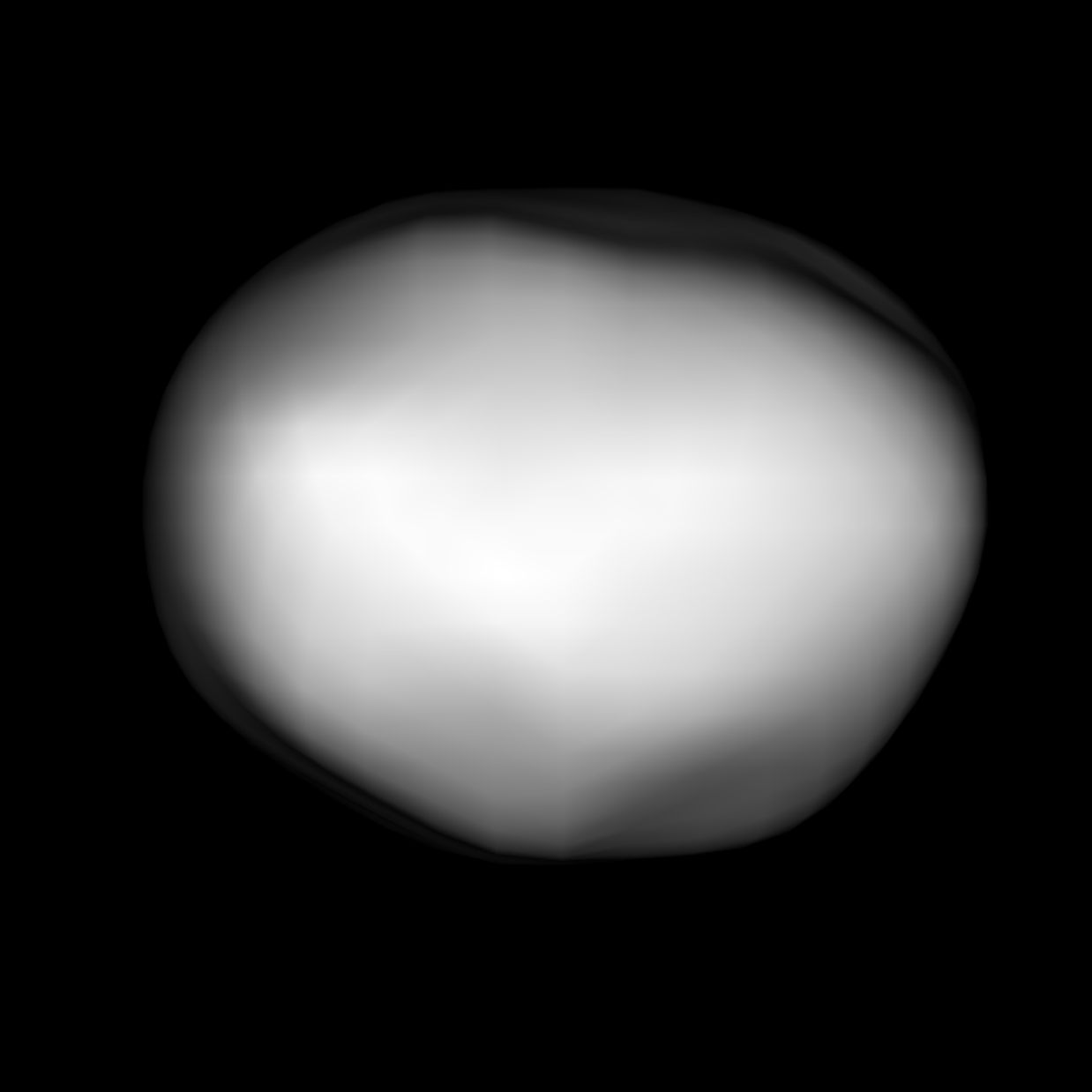}
    \hspace{.005\linewidth}
    \includegraphics[width=.3\linewidth]{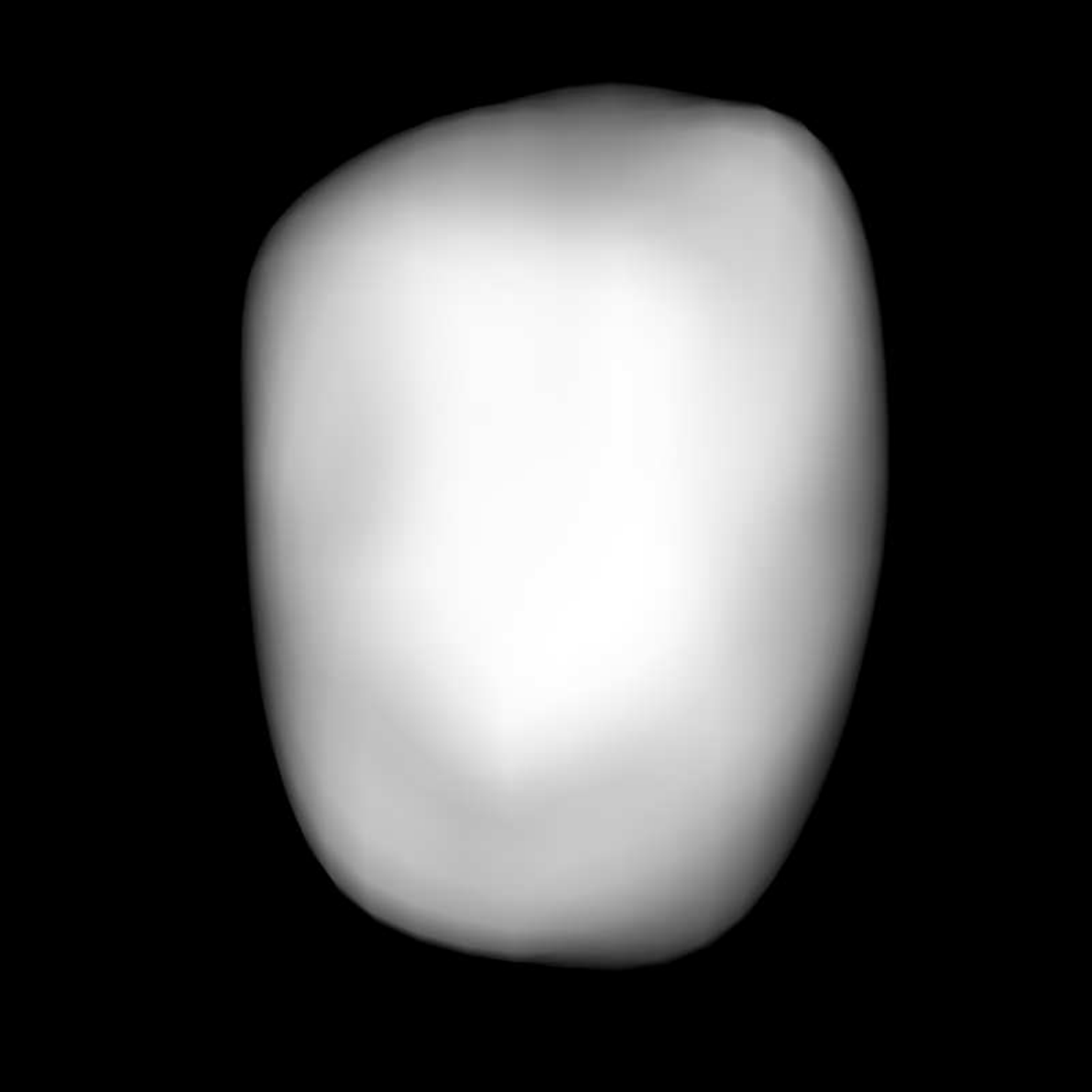}
    }
    \\
    \resizebox{\hsize}{!}{
    \includegraphics[width=.3\linewidth]{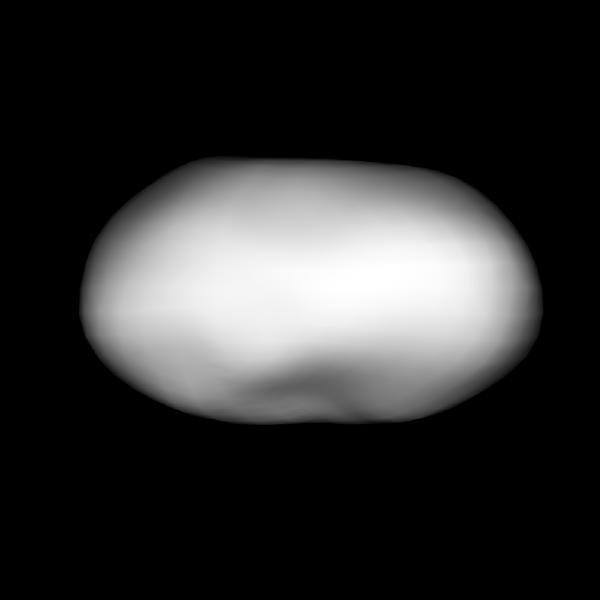}
    \hspace{.005\linewidth}
    \includegraphics[width=.3\linewidth]{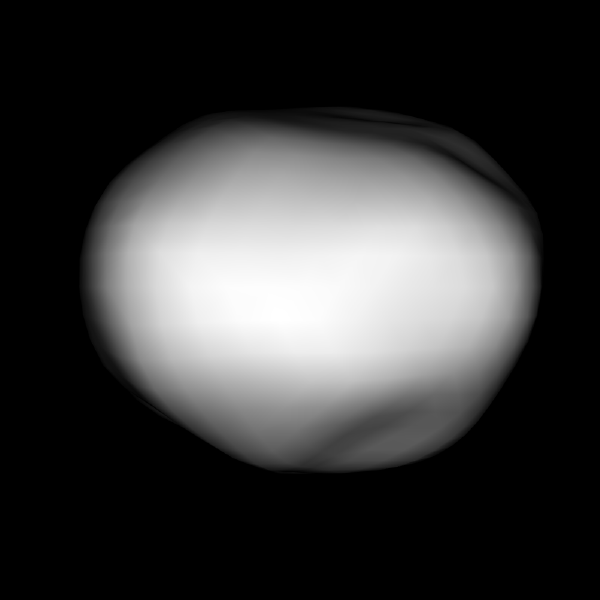}
    \hspace{.005\linewidth}
    \includegraphics[width=.3\linewidth]{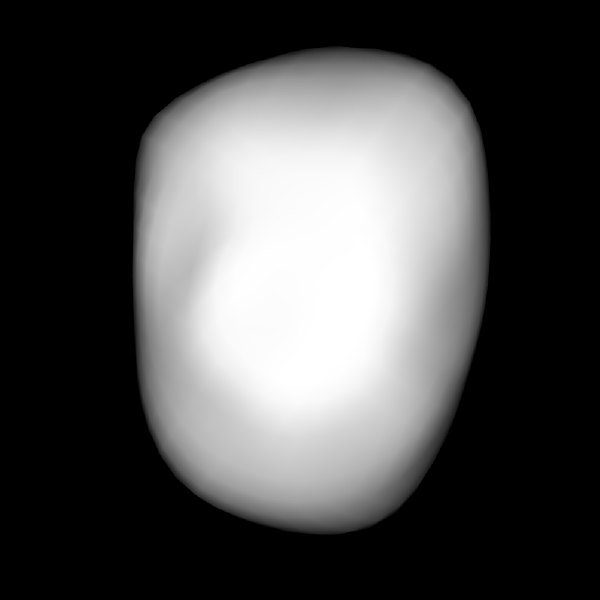}
    }
  \caption{The best-fit shape model (\emph{top}) and the alternative shape model (\emph{bottom}) from three different viewing geometries. The first two are \emph{equator-on} views rotated by 90$^{\circ}$ and the third one is a \emph{pole-on} view. The model is lit up artificially to highlight its finer surface details.
  }
  \label{fig:model}
\end{figure}

In Fig.~\ref{fig:model} and Table~\ref{tab:model}, we present the best-fit and alternative shape models and their parameters. The best-fit model is a revised version of the shape model published in \citet{vernazza2021vlt}, based on a larger dataset. Its pole coordinates were allowed to converge along with the shape to best-fit the data. The alternative model is a shape model that converged with the pole fixed according to the coordinates inferred from our orbital model (Sect.~\ref{sec:orbit}).

\begin{figure}
  \resizebox{\hsize}{!}{
    \includegraphics[width=\textwidth]{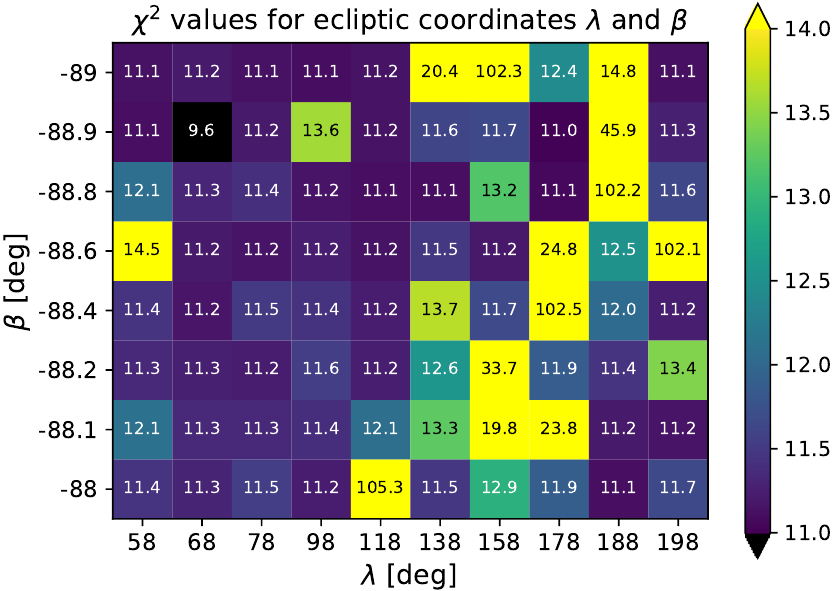}
    }
  \caption{
    $\chi^2$ map for pole coordinates $\lambda$ and $\beta$, where each value of $\chi^2$ is represented by a colour from the colour bar. The \emph{black} tiles indicate values below the range of the colour bar, while the \emph{yellow} tiles indicate values above the range of the colour bar.
  }
  \label{fig:chi2_map_ADAM}
\end{figure}

The $\chi^2$ map in Fig.~\ref{fig:chi2_map_ADAM} shows why the pole solution from ADAM remains unconstrained. Due to the close absolute distance of coordinates, there are many viable models. The best-fit model, denoted by a black tile, is in the upper left area of the map, while the alternative model is in a different local minimum in the bottom right area of the map. We propose that this alternative local minimum is where the true orbital pole lies. In Sect.~\ref{sec:orbit}, we even present a counterexample model based on the shape pole solution, to support this.

\begin{figure}
  \resizebox{\hsize}{!}{
    \includegraphics[width=\textwidth]{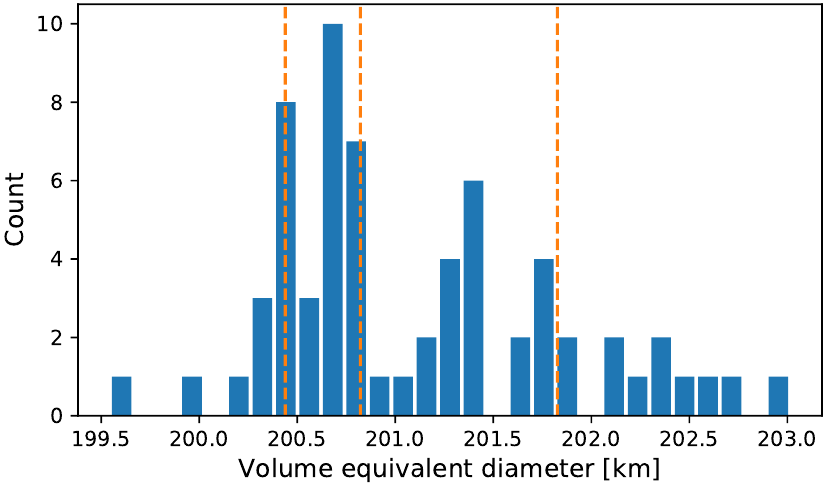}
    }
  \caption{
    Distribution of $D_{\rm{eq}}$ values based on shape models with $\chi^2$ values below 12. The 16\%, 50\%, and 84\% percentiles of the distribution are denoted by \emph{orange} lines.
  }
  \label{fig:volume_spread}
\end{figure}

We plotted a distribution of volume values from the dataset of models obtained from the $\chi^2$ mapping (Figure~\ref{fig:volume_spread}). By looking at the 16\%, 50\%, and 84\% percentiles of the distribution we determined the uncertainties of $D_{\rm{eq}}$ in Table~\ref{tab:model}.

\begin{figure*}
\centering
\includegraphics[width=.15\linewidth]{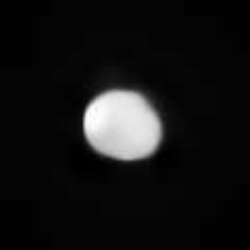}
\hspace{.005\linewidth}
\includegraphics[width=.15\linewidth]{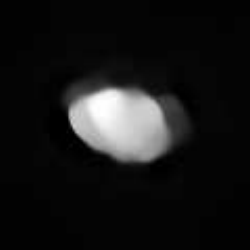}
\hspace{.005\linewidth}
\includegraphics[width=.15\linewidth]{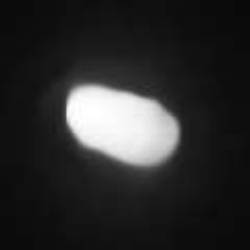}
\hspace{.005\linewidth}
\includegraphics[width=.15\linewidth]{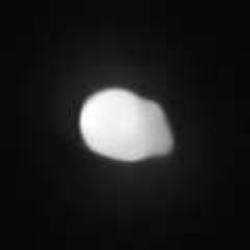}
\hspace{.005\linewidth}
\includegraphics[width=.15\linewidth]{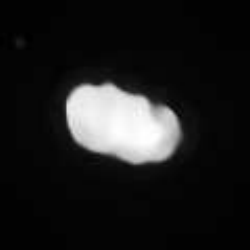}
\hspace{.005\linewidth}
\includegraphics[width=.15\linewidth]{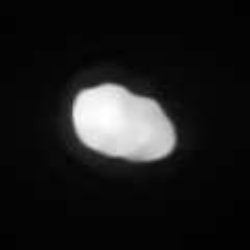}
\\
\includegraphics[width=.15\linewidth]{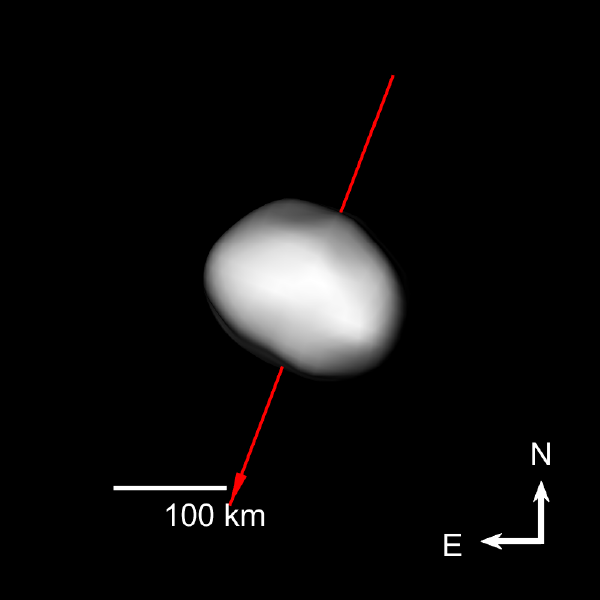}
\hspace{.005\linewidth}
\includegraphics[width=.15\linewidth]{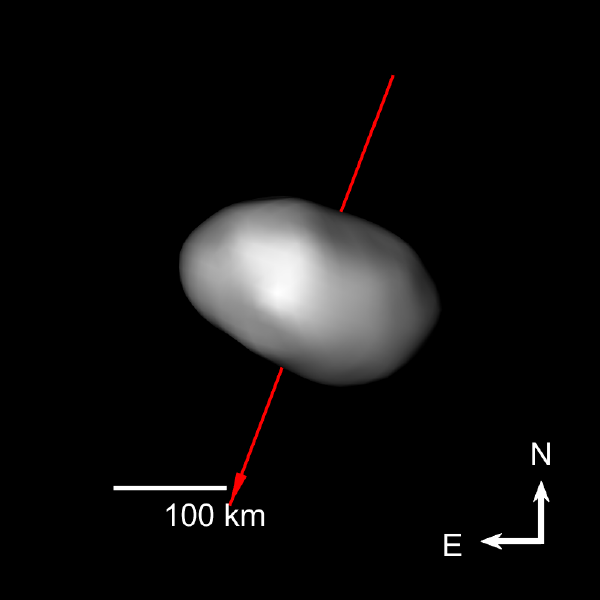}
\hspace{.005\linewidth}
\includegraphics[width=.15\linewidth]{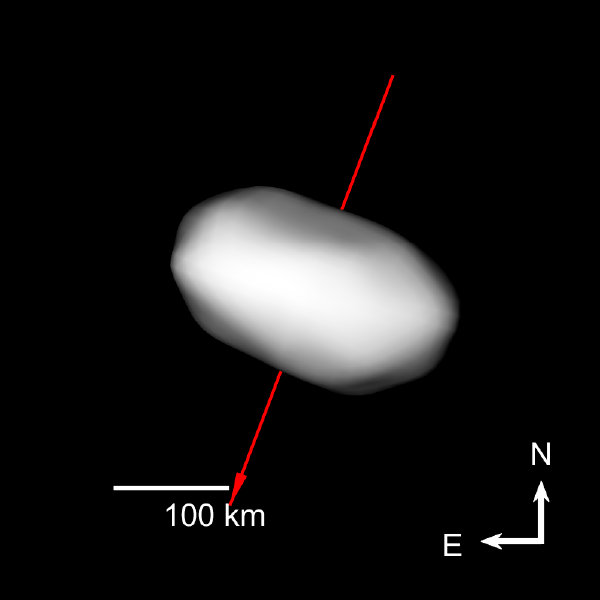}
\hspace{.005\linewidth}
\includegraphics[width=.15\linewidth]{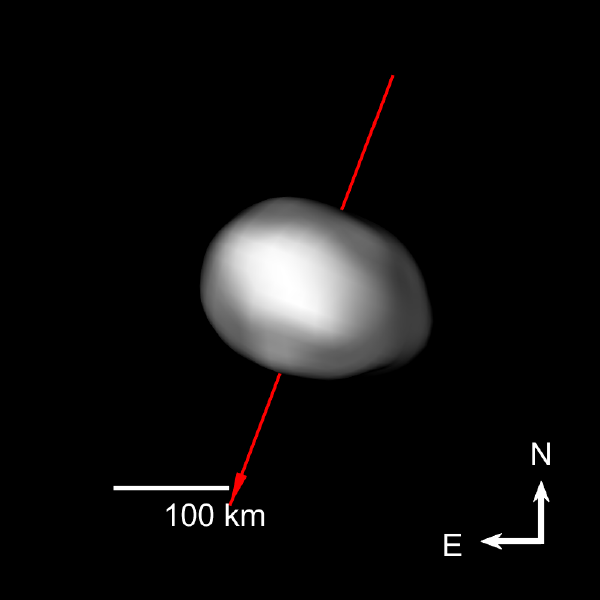}
\hspace{.005\linewidth}
\includegraphics[width=.15\linewidth]{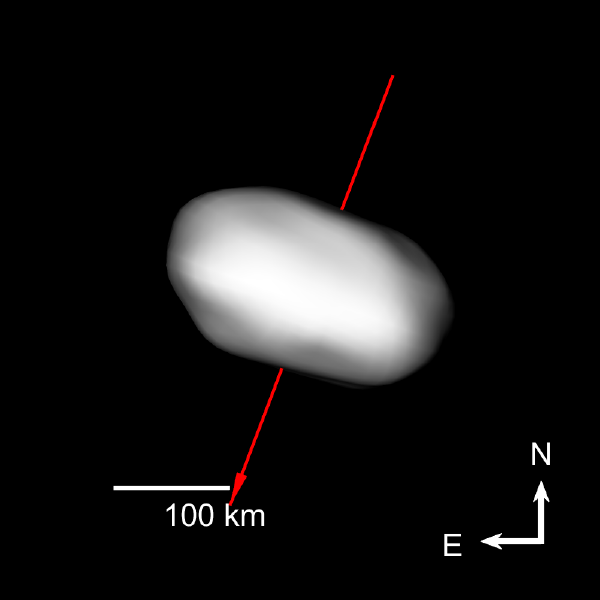}
\hspace{.005\linewidth}
\includegraphics[width=.15\linewidth]{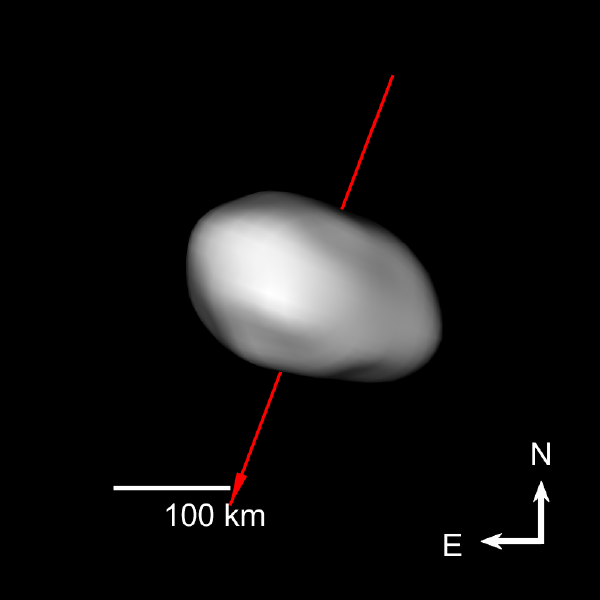}
\\
\includegraphics[width=.15\linewidth]{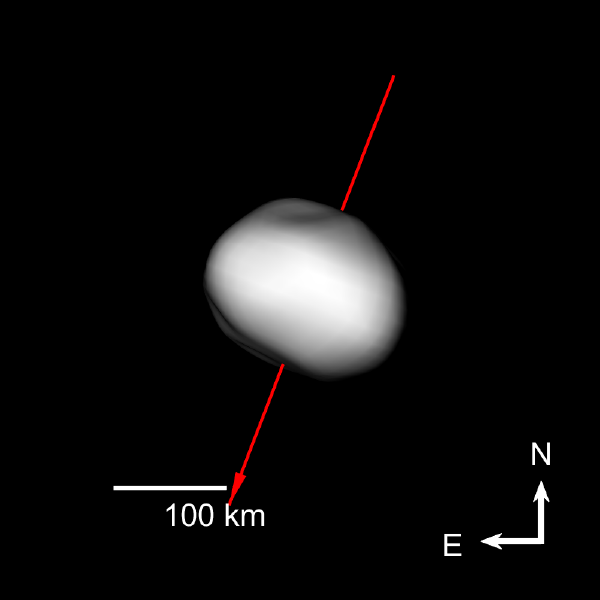}
\hspace{.005\linewidth}
\includegraphics[width=.15\linewidth]{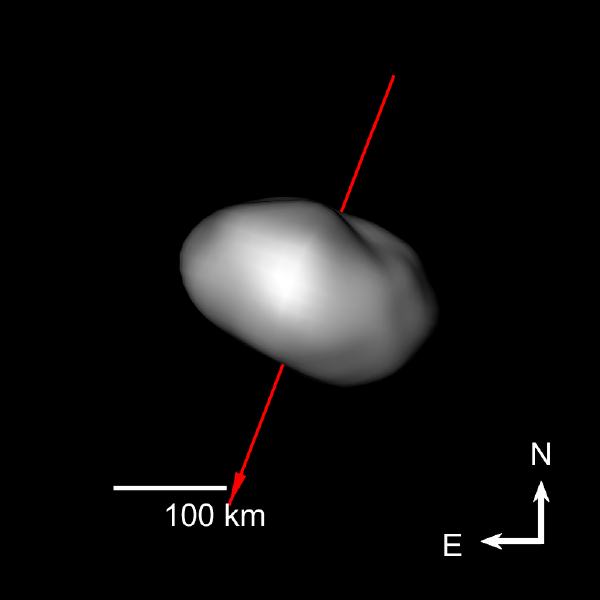}
\hspace{.005\linewidth}
\includegraphics[width=.15\linewidth]{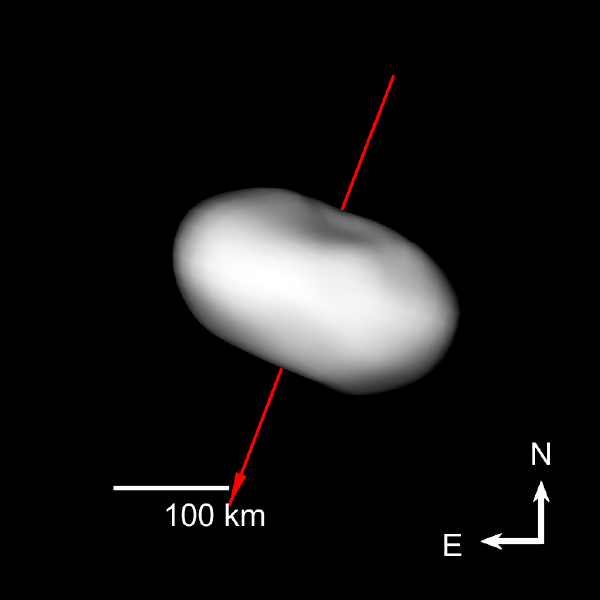}
\hspace{.005\linewidth}
\includegraphics[width=.15\linewidth]{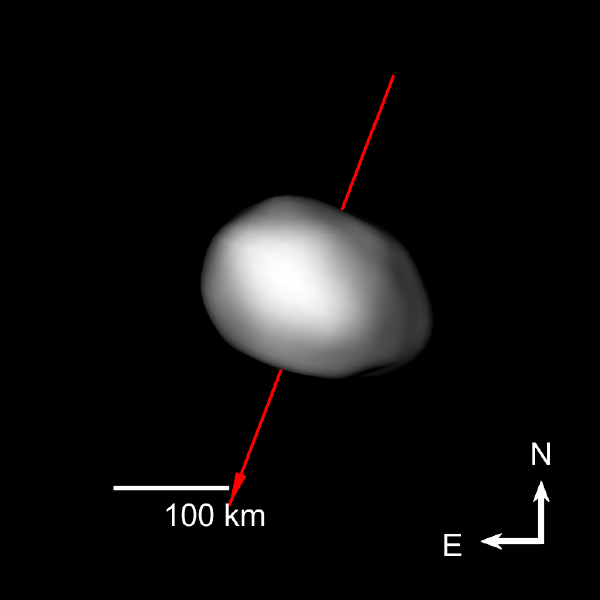}
\hspace{.005\linewidth}
\includegraphics[width=.15\linewidth]{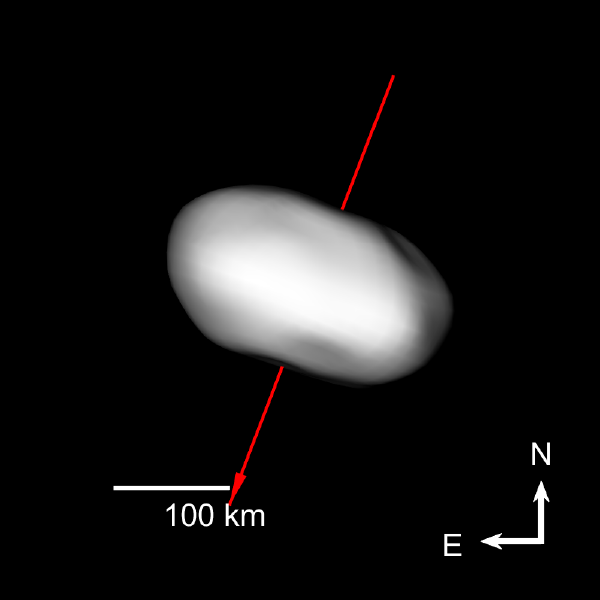}
\hspace{.005\linewidth}
\includegraphics[width=.15\linewidth]{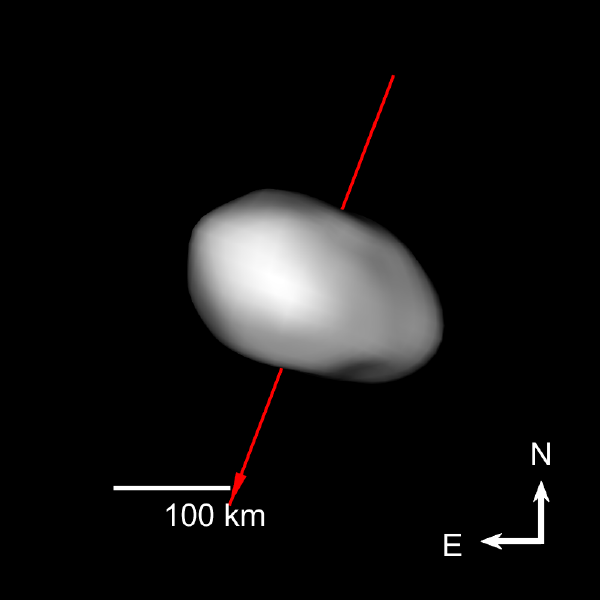}
\caption{
AO images of Elektra (\emph{top}), the best-fit shape model (\emph{middle}) and the alternative shape model (\emph{bottom}),
shown from the same viewing angle.
The rotation axis is indicated by the red arrow.
}
\label{fig:comparison}
\end{figure*}

The comparison of these two models with several AO images from the 2019 data set is given in Fig.~\ref{fig:comparison}. In the comparison and Fig.~\ref{fig:model} we see that the alternative model's overall shape is more rounded with fewer surface features in some areas. However, this is in the range of expected differences dependent on regularizations. Both models are statistically and visually almost identical.

In the last two columns of Fig.~\ref{fig:comparison}, we see that the local topography, present in the two AO images, is not captured by either of the models. We tried using the subdivision shape representation and locally forcing those features. But, it seems that they are not supported by the rest of the dataset, and forcing them just makes the overall fit worse.  

For both models, we used the following weights: $w_{\rm LC} = 1.5$,
$w_{\rm AO} = 0.85$, $w_{\rm OC} = 0.019$ and standard values for the
regularization weights.

In Appendix \ref{app:lightcurves}, we compare the best-fit model to all the observed lightcurves.

\begin{table}
\caption{Parameters of the best-fit (\emph{first} line) and alternative (\emph{second} line) models.}
\label{tab:model}
\centering
\footnotesize
\begin{tabular}{*{5}{c}}
\hline\hline
$\lambda$ & $\beta$ & $P$ & $a \times b \times c$ & $D_{\rm{eq}}$ \\
{[deg]} & {[deg]} & {[h]} & {[km]} & {[km]} \\
\hline
$68.5$ & $-88.9$ & $5.224663$ & $267 \times 202 \times 151$ & $201.4 \pm 1$ \\
$188.2$ & $-88.1$ & $5.224664$ & $273 \times 230 \times 151$ & $202.3 \pm 2$ \\
\hline
\end{tabular}
\tablefoot{
$\lambda$, $\beta$ denote the ecliptic coordinates of the pole,
$P$ the period of rotation,
$a$, $b$, $c$ the extents along the main axes~\footnote{The extents $a$, $b$, $c$ were obtained by the overall dimensions technique \citep{torppa2008asteroid}.},
$D_{\textrm{eq}}$ the diameter of the volume-equivalent sphere.
The 1-$\sigma$ uncertainties of $D_{\textrm{eq}}$ are based on the distribution from Fig.~\ref{fig:volume_spread}.}

\end{table}

\begin{figure*}
\sidecaption
   \includegraphics[width=12cm]{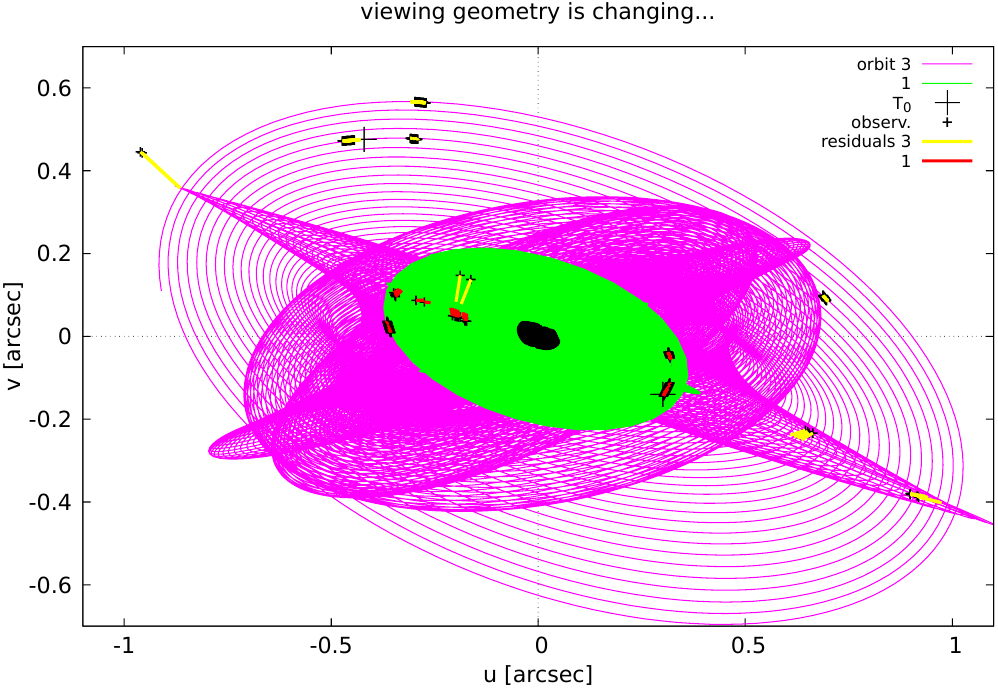}
     \caption{
     Keplerian model of orbits 1 and 3 with $\chi^2 = 2195$ over the time-span of 1707 days.
     The orbits are plotted in the (\emph{u; v}) coordinates (\emph{green} and \emph{pink} lines, respectively),
     together with observed positions (\emph{black crosses}),
     and residuals (\emph{red} and \emph{yellow} lines, respectively).
     Elektra’s shape for one of the epochs is overplotted in \emph{black}.
     }
     \label{fig:graph_of_model_Kepler}
\end{figure*}


\subsection{Multipole coefficients}

Our orbital model has to be evolved over approximately 1000 days.
On such a time scale, approximating the central body as a point mass is insufficient.
Therefore, we computed a multipole expansion of the gravitational field,
up to the order of $\ell = 10$.
The expansion is based on the Delaunay triangulation of the ADAM shape model, made using the TetGen program \citep{TetGen}.
The respective definitions of the multipole coefficients are as follows:
\begin{equation}
    \label{eq:Cl0}
    C_{\ell 0} = \frac{1}{MR^{\ell}} \int_{V}^{} \rho |\boldsymbol{r}|^{\ell} P_{\ell} (\cos{\theta}) \,\text{d}V\,,
\end{equation}
\begin{equation}
    \label{eq:Clm}
    C_{\ell m} = \frac{2}{MR^{\ell}} \frac{(\ell - m)!}{(\ell + m)!} \int_{V}^{} \rho |\boldsymbol{r}|^{\ell} P_{\ell m} (\cos{\theta}) \cos{(m \phi)} \,\text{d}V\,,
\end{equation}
\begin{equation}
    \label{eq:Slm}
    S_{\ell m} = \frac{2}{MR^{\ell}} \frac{(\ell - m)!}{(\ell + m)!} \int_{V}^{} \rho |\boldsymbol{r}|^{\ell} P_{\ell m} (\cos{\theta}) \sin{(m \phi)} \,\text{d}V\,,
\end{equation}
where $r$, $\theta$, $\phi$ are the body-frozen spherical coordinates,
$M$ the mass of the body,
$R$ the reference radius of the gravitational model,
$P_{\ell}$, $P_{\ell m}$ the Legendre and associated Legendre polynomials, and
$C_{\ell m}$, $S_{\ell m}$ the respective coefficients.

In Table~\ref{tab:multipole_coef} we list the results of Eqs.~(\ref{eq:Cl0}--\ref{eq:Slm}) evaluated for our best-fit shape model with homogeneous density. It turned out that an expansion up to the order $\ell = 2$ is sufficient enough to account for the precession of orbits and the higher orders (up to $\ell = 6$) were used to confirm that their contribution is in fact negligible.

Most notable is the oblateness coefficient $J_2 = -C_{20} \approx 0.159$.
This is a relatively high value;
only a few 100-km asteroids have a similar $c/a$ axial ratio as (130) Elektra \citep{vernazza2021vlt},
for instance
(7)~Iris,
(16)~Psyche,
(22)~Kalliope, or
(45)~Eugenia.
The value inferred from shape will be compared to the dynamical oblateness in Sect.~\ref{sec:C20}.

The alternative shape model has an almost equivalent $J_2 \simeq 0.162$.
This corresponds to an uncertainty of only $0.003$.
We verified that the discretisation error of $J_2$,
arising from the finite number of tetrahedrons,
is less than $0.001$.

\begin{table}
\caption{Multipole coefficients (up to $\ell = 6$) of Elektra’s gravitational field
derived from the ADAM shape, assuming homogeneous density.
}
\label{tab:multipole_coef}
\centering\footnotesize\sf
\begin{tabular}{*{4}{l}}
\hline\hline
$C_{00}$ & $+1.00000000$                &          &                              \\

$C_{10}$ & $+0.00000000$                &          &                              \\
$C_{11}$ & $+0.00000000$                & $S_{11}$ & $+0.00000000$                \\

$C_{20}$ & $-1.59109015 \times 10^{-1}$ &          &                              \\
$C_{21}$ & $-7.06376507 \times 10^{-5}$ & $S_{21}$ & $-1.73622001 \times 10^{-4}$ \\
$C_{22}$ & $-4.46764263 \times 10^{-2}$ & $S_{22}$ & $+5.16249277 \times 10^{-6}$ \\

$C_{30}$ & $+1.52377403 \times 10^{-3}$ &          &                              \\
$C_{31}$ & $-4.71344916 \times 10^{-4}$ & $S_{31}$ & $-2.13299430 \times 10^{-3}$ \\
$C_{32}$ & $+1.68475688 \times 10^{-3}$ & $S_{32}$ & $+7.99513867 \times 10^{-4}$ \\
$C_{33}$ & $-7.99140262 \times 10^{-4}$ & $S_{33}$ & $+1.61094232 \times 10^{-3}$ \\

$C_{40}$ & $+5.47386330 \times 10^{-2}$ &          &                              \\
$C_{41}$ & $+1.84595619 \times 10^{-3}$ & $S_{41}$ & $-1.98090953 \times 10^{-3}$ \\
$C_{42}$ & $+4.18945086 \times 10^{-3}$ & $S_{42}$ & $-1.56934242 \times 10^{-4}$ \\
$C_{43}$ & $+1.63489342 \times 10^{-4}$ & $S_{43}$ & $+8.16971619 \times 10^{-6}$ \\
$C_{44}$ & $+1.19047276 \times 10^{-4}$ & $S_{44}$ & $-1.04901431 \times 10^{-4}$ \\

$C_{50}$ & $-3.05649267 \times 10^{-3}$ &          &                              \\
$C_{51}$ & $+3.70902439 \times 10^{-4}$ & $S_{51}$ & $+1.27208068 \times 10^{-4}$ \\
$C_{52}$ & $-5.69391535 \times 10^{-4}$ & $S_{52}$ & $-2.33343309 \times 10^{-4}$ \\
$C_{53}$ & $+7.24854310 \times 10^{-5}$ & $S_{53}$ & $-1.50240727 \times 10^{-4}$ \\
$C_{54}$ & $-3.17247273 \times 10^{-5}$ & $S_{54}$ & $-1.43162546 \times 10^{-5}$ \\
$C_{55}$ & $+1.64621165 \times 10^{-5}$ & $S_{55}$ & $-1.89458984 \times 10^{-5}$ \\

$C_{60}$ & $-2.54715420 \times 10^{-2}$ &          &                              \\
$C_{61}$ & $-1.11217188 \times 10^{-3}$ & $S_{61}$ & $+1.69598512 \times 10^{-3}$ \\
$C_{62}$ & $-9.25476914 \times 10^{-4}$ & $S_{62}$ & $+6.44072903 \times 10^{-5}$ \\
$C_{63}$ & $-4.83382376 \times 10^{-5}$ & $S_{63}$ & $+1.27210479 \times 10^{-5}$ \\
$C_{64}$ & $-6.40791927 \times 10^{-6}$ & $S_{64}$ & $+6.56867385 \times 10^{-6}$ \\
$C_{65}$ & $-2.07536310 \times 10^{-6}$ & $S_{65}$ & $-7.77372686 \times 10^{-7}$ \\
$C_{66}$ & $+6.05696810 \times 10^{-7}$ & $S_{66}$ & $+5.52013261 \times 10^{-7}$ \\
\hline
\end{tabular}
\end{table}

\section{Orbital model}\label{sec:orbit}

\subsection{The dynamics included in the model}

We use the \emph{N}-body model called {\tt Xitau}%
\footnote{\url{https://sirrah.troja.mff.cuni.cz/~mira/xitau/}},
based on the Bulirsch-Stoer integrator from
\cite{Levison_1994Icar..108...18L}
that was substantially modified
\citep{brovz2017advanced,brovz2021advanced,Broz_2022A&A...657A..76B}
to include
the multipole expansion (up to the order $\ell$ = 10),
the tidal evolution,
the external tides, and in particular, the fitting `machinery'.
To compare the model with the observations,
we use the following unreduced metric \footnote{i.e. not divided by N; the number of measurements}:

\begin{equation}
    \label{eq:metric}
    \chi^2 = \chi_{\text{sky}}^2 + \chi_{\text{sky2}}^2 + w_{\rm ao} \, \chi_{\text{ao}}^2\,,
\end{equation}

\begin{equation}
    \label{eq:metric_sky}
    \chi_{\text{sky}}^2 = \sum_{j=1}^{N_{\text{bod}}} \sum_{i=1}^{N_{\text{sky}}} \Biggr[ \frac{(\Delta u_{ji})^2}{\sigma_{\text{sky major}\,ji}^2} + \frac{(\Delta v_{ji})^2}{\sigma_{\text{sky minor}\,ji}^2} \Biggr]\,,
\end{equation}

\begin{equation}
    \label{eq:metric_sky2}
    \chi_{\text{sky2}}^2 = \sum_{i=1}^{N_{\text{sky2}}} \Biggr[ \frac{(\Delta u_{i})^2}{\sigma_{\text{sky major}\,i}^2} + \frac{(\Delta v_{i})^2}{\sigma_{\text{sky minor}\,i}^2} \Biggr]\,,
\end{equation}

\begin{equation}
    \label{eq:metric_ao}
    \chi_{\text{ao}}^2 = \sum_{i=1}^{N_{\text{ao}}} \sum_{k=1}^{360} \frac{(u_{ik}' - u_{ik})^2 (v_{ik}' - v_{ik})^2}{\sigma_{\text{ao}\,i}^2} \,,
\end{equation}
where the index $i$ corresponds to the observational data,
$j$ to individual bodies,
$k$ to angular steps of silhouette data,
$'$~to synthetic data interpolated to the times of observations,
$u$, $v$ are the sky-plane coordinates, and
$\sigma$ the observational uncertainties along the two axes
(denoted as ‘major’ and ‘minor’ for ellipsoidal uncertainties).

The 'goodness-of-fit' terms $\chi_{\text{sky}}^2$ and $\chi_{\text{sky2}}^2$
correspond to the absolute astrometry (i.e. with respect to body~0) and
the relative astrometry (e.g. body 2 with respect to body 1).
Optionally, we also include the regularisation term $\chi^2_{\rm ao}$,
derived from silhouettes (as explained in \citealt{brovz2021advanced}),
which prevents pole orientations incompatible with the shape. 
This term is multiplied by the weight $w_{\rm ao}$,
which bears no connection to the weight term $w_{\rm ao}$ from Eq.~(\ref{eq:ADAM_metric}).  

Using both $\chi_{\text{sky}}^2$ and $\chi_{\text{sky2}}^2$ is useful because they are not exactly the same measurements. Computing the relative positions removes any systematics related to the photocentre and provides more precise (relative) information. Note that this rapidly increases the overall number of measurements, as each AO image where two or three moons are visible at the same time is counted again for each relative measurement taken.

In the following, we present two orbital models of increasing complexity.
In both of them, we account for the external tide exerted by the Sun.
The necessary ephemerides were obtained from the Jet Propulsion Laboratory
(JPL; \citealt{Park_2021AJ....161..105P}).


\subsection{Keplerian model}

Initially, by manual fitting and then by converging with the simplex algorithm \citep{nelder1965simplex},
we constructed a simplified Keplerian model for the already well-known moons: S/2014 (130) and S/2003 (130), that is, a model of orbits 1 and 3.
The masses of the moons were neglected and the central body was taken as a point of mass. 
These preliminary orbits are shown in Fig.~\ref{fig:graph_of_model_Kepler}.

In the case of the Elektra system, a Keplerian model is accurate enough for a time span of about a month,
as can be seen in \citet{Berdeu_2022A&A...658L...4B},
but is insufficient for the time span of 1707 days.
The resulting value of $\chi^2 = 2195$ is too high compared to the number of measurements (i.e. 430).
The residuals in Fig.~\ref{fig:graph_of_model_Kepler} exhibit large systematics.
Especially in the top left part, the synthetic orbit is nowhere close to the observations.
This is due to the missing nodal precession, which must be present
given the non-negligible oblateness ($J_2 = -C_{20}$) of the central body.


\subsection{Best-fit quadrupole model}

\begin{figure*}
\sidecaption
   \includegraphics[width=12cm]{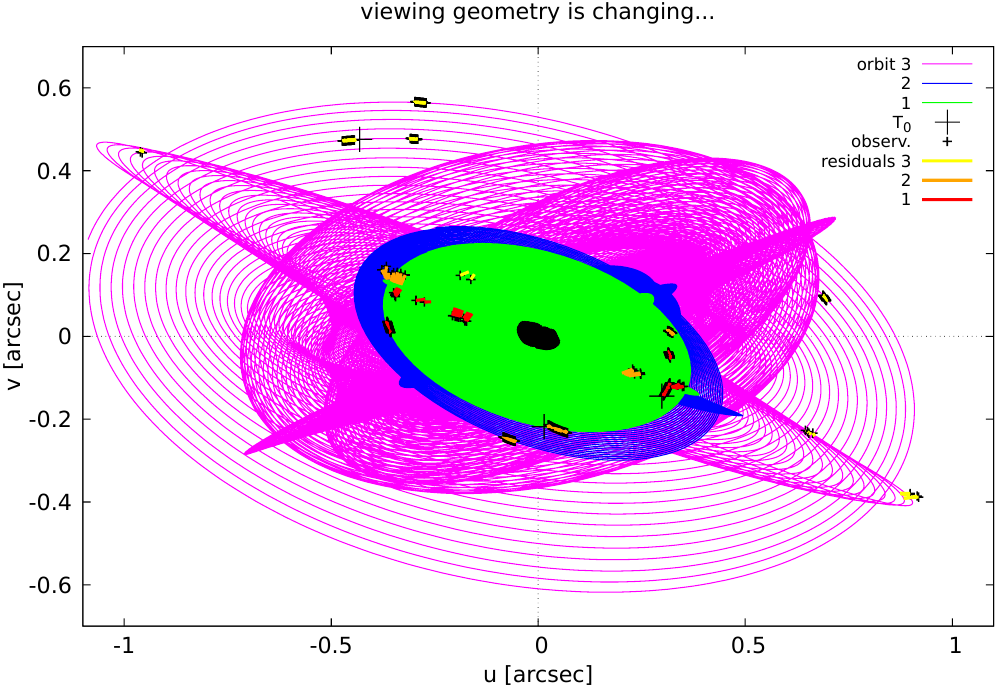}
     \caption{
     Quadrupole model of the orbits 1, 2 and 3 with $\chi^2$ = 1084 over the time-span of 1707 days.
      The orbits are plotted in the (\emph{u; v}) coordinates (\emph{green}, \emph{blue} and \emph{pink} lines, respectively),
      together with observed positions (\emph{black crosses}),
     and residuals (\emph{red}, \emph{orange} and \emph{yellow} lines, respectively).
     Elektra’s shape for one of the epochs is overplotted in \emph{black}.
     }
     \label{fig:graph_of_model_default}
\end{figure*}

\begin{figure}
  \resizebox{\hsize}{!}{
    \includegraphics[width=\textwidth]{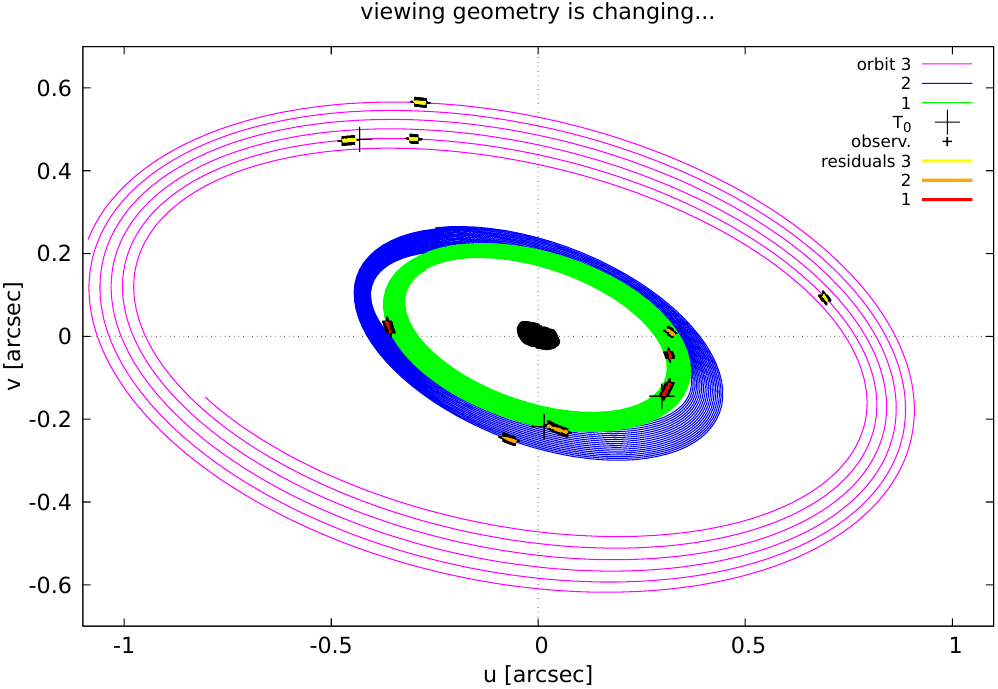}
    }
  \resizebox{\hsize}{!}{
    \includegraphics[width=\textwidth]{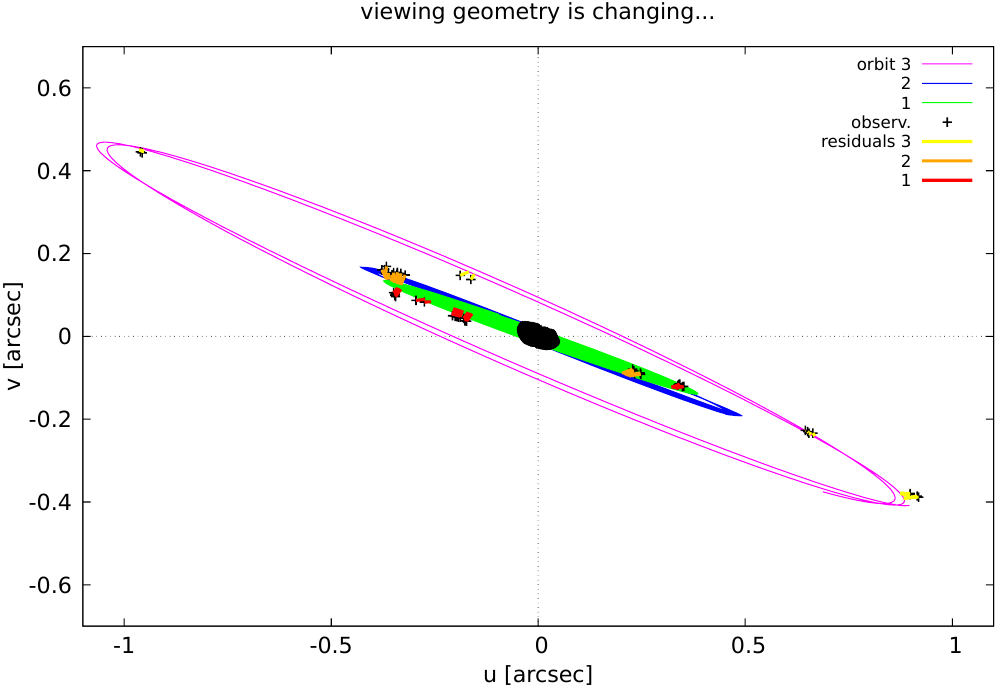}
    }
  \caption{
  Same as Fig.~\ref{fig:graph_of_model_default},
  but plotted separately for each data set:
  \citet{Berdeu_2022A&A...658L...4B} (\emph{top}) and \citet{vernazza2021vlt} (\emph{bottom}).
  }
  \label{fig:graph_of_model_separate}
\end{figure}

We present a quadrupole model of the full system in Figs.~\ref{fig:graph_of_model_default}~and~\ref{fig:graph_of_model_separate}.
In this more complex model, the masses of the moons are taken into account
and the gravitational field of the central body was expanded up to the order $\ell$ = 2
(according to Table~\ref{tab:multipole_coef}), assuming homogeneous density.
When evaluated with the full metric (Eq.~(\ref{eq:metric})) with the weight $w_{\rm ao} = 0.3$, the model is a best-fit one with $\chi^2 = 1084$.

This model is the result of a long series of models, which were being improved with every iteration. It was much more efficient to
only use up to $\ell$ = 2, which truly is enough for a satisfying model.

\begin{figure}
  \resizebox{\hsize}{!}{
    \includegraphics[width=\textwidth]{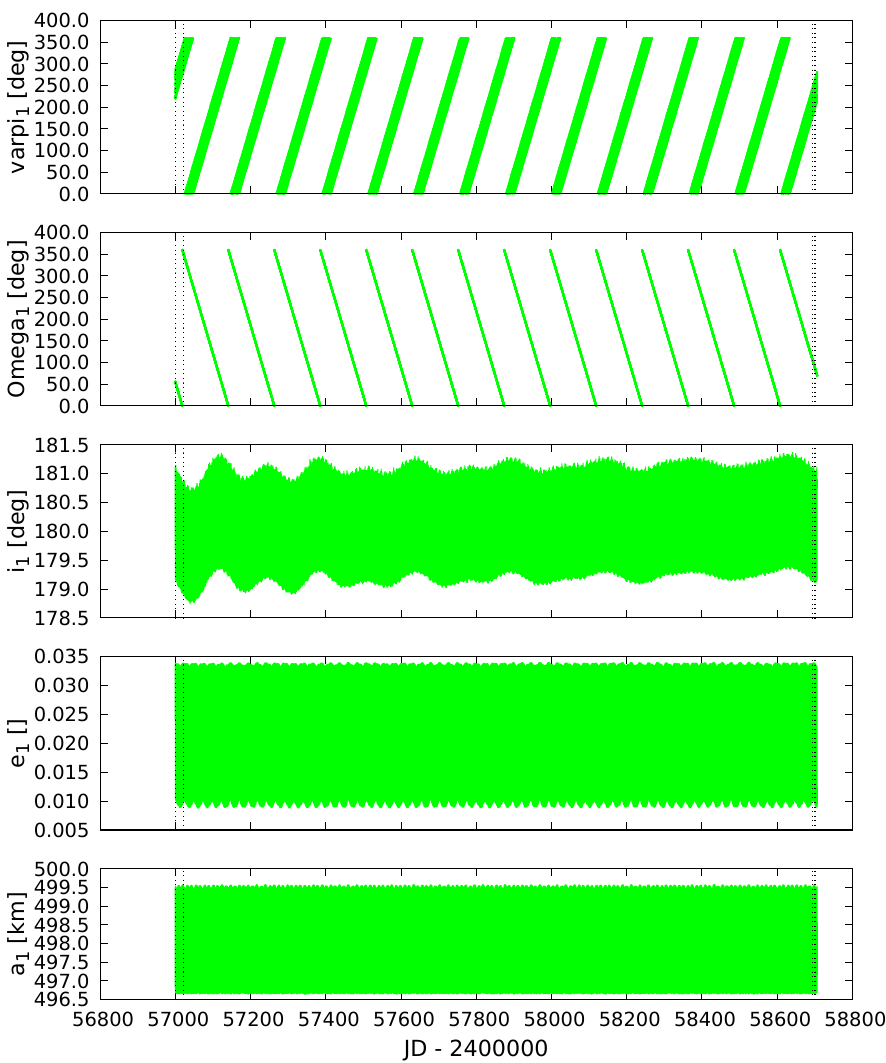}
    }
  \caption{
     Evolution of the osculating elements of the inner moon S/2014 1 plotted over the time span of 1707 days.
     Shown is the semimajor axis $a_1$,
     eccentricity $e_1$,
     inclination $i_1$,
     the longitude of the ascending node $\Omega_1$, and
     the longitude of the periapsis $\varpi_1$.
  }
  \label{fig:graph_of_osculating_elements_1}
\end{figure}

\begin{figure}
  \resizebox{\hsize}{!}{
    \includegraphics[width=\textwidth]{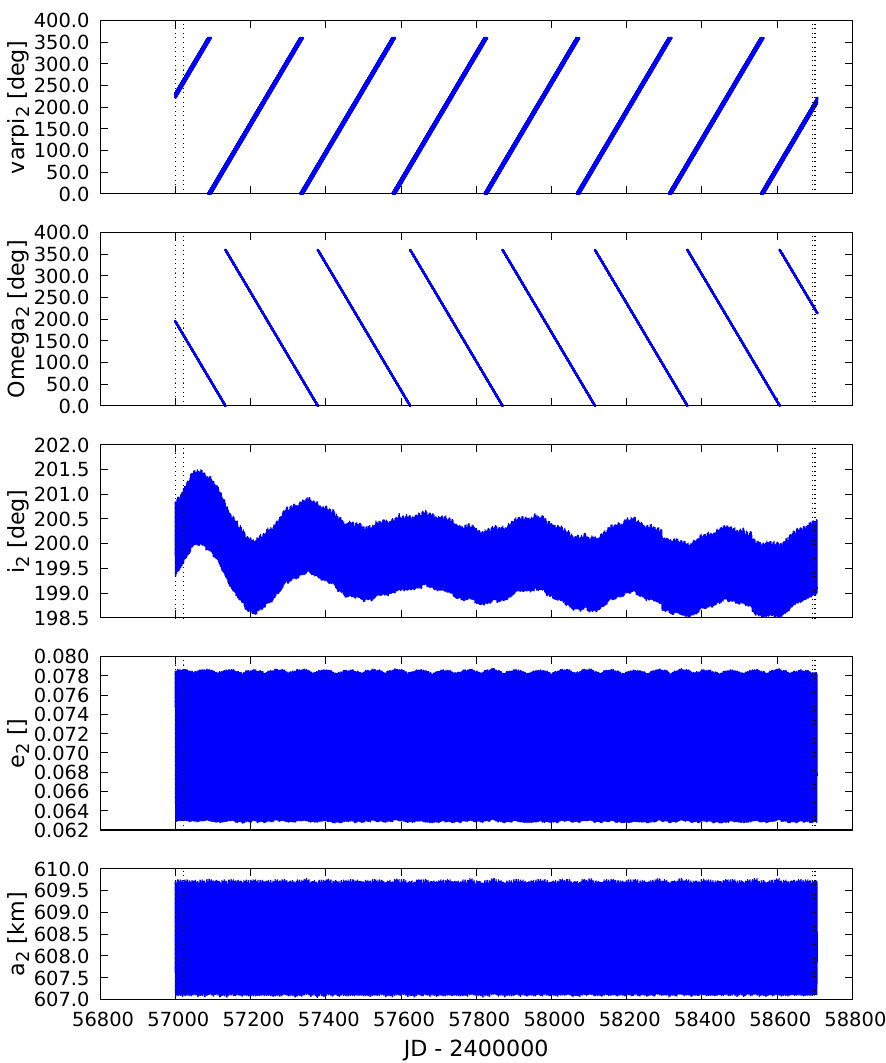}
    }
  \caption{
Same as Fig.~\ref{fig:graph_of_osculating_elements_1},
     but for the middle moon S/2014 2.}
  \label{fig:graph_of_osculating_elements_2}
\end{figure}

\begin{figure}
  \resizebox{\hsize}{!}{
    \includegraphics[width=\textwidth]{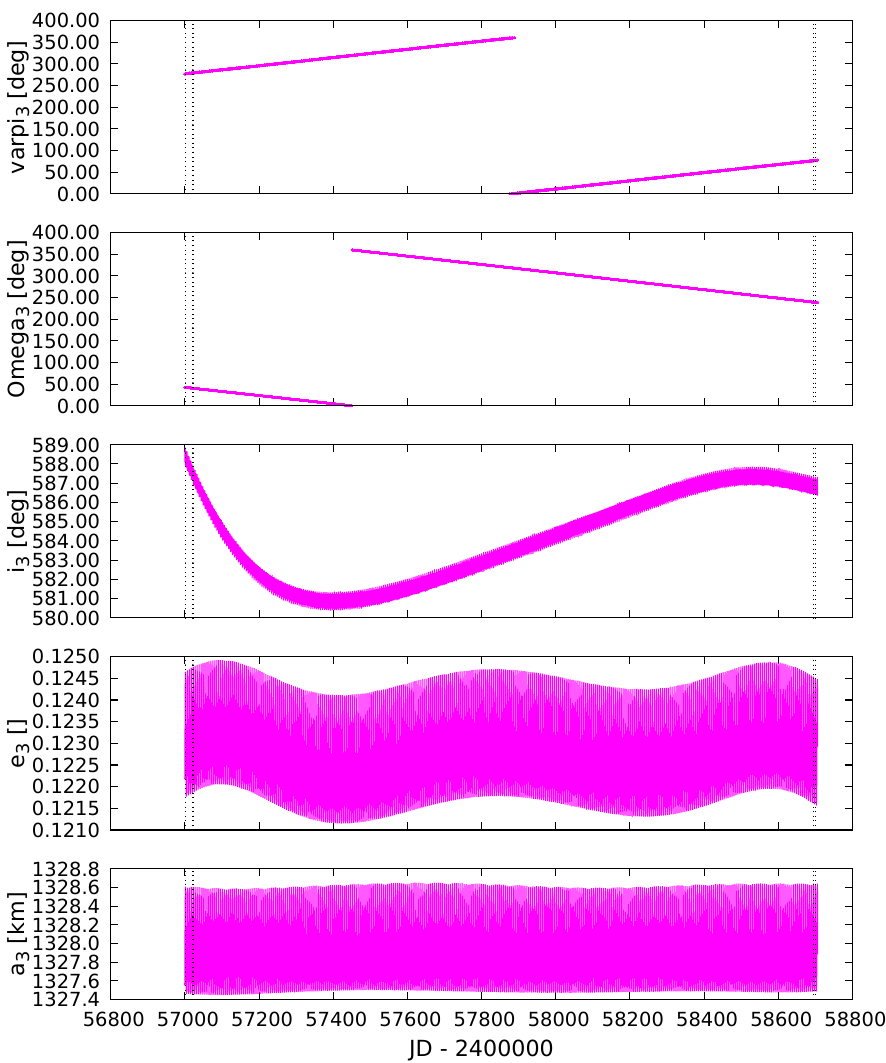}
    }
  \caption{
Same as Fig.~\ref{fig:graph_of_osculating_elements_1},
     but for the outer moon S/2003.}
  \label{fig:graph_of_osculating_elements_3}
\end{figure}

We list the parameters of this best-fit model in Tables~\ref{tab:parameters} and \ref{tab:dependent_parameters}.
They are given as osculating for the epoch $T_0 = 2457021.567880$ (TDB),
but these elements are not constant.
Due to the oblateness of Elektra and massive moons, 
the orbital elements oscillate over time,
as can be seen in Figs.~\ref{fig:graph_of_osculating_elements_1}~--~\ref{fig:graph_of_osculating_elements_3}.
The peak-to-peak amplitudes of some of these elements are substantial,
so it's no wonder the Keplerian model was insufficient.
From the circulation of the arguments of pericentre,
we determined the apsidal precession of the moons:
$\dot\omega_1 = 5.9 \,{\rm deg\,day}^{-1}$,
$\dot\omega_2 = 2.9 \,{\rm deg\,day}^{-1}$, and
$\dot\omega_3 = 0.19 \,{\rm deg\,day}^{-1}$.

From the precession cycles of $\Omega$ \& $\varpi$ (Figs.~\ref{fig:graph_of_osculating_elements_1}~--~\ref{fig:graph_of_osculating_elements_3}) follows the uniqueness of our solution.
The two innermost orbits have many cycles, about 14 cycles of $\Omega_1$ \& $\varpi_1$ and about 7 cycles of $\Omega_2$ \& $\varpi_2$. 
A correct orbital solution could be one with two and one cycle less. 
Thankfully we also have the outer moon and its orbit with only a half cycle of $\Omega_3$ \& $\varpi_3$, which just cannot be made one cycle more or less, without being a completely different and unviable solution. Thus, even the number of the two innermost moons' precession cycles is well constrained because the central body is the same.         

Having the precise mass of Elektra, $(6.606 \substack{+0.007 \\ -0.013}) \times 10^{18}\,{\rm kg}$,
we can combine it with the volume of the shape model,
$(4.3 \pm~0.1) \times 10^{6}\,{\rm km}^{3}$,
and obtain the precise bulk density,
$\bar{\rho} = (1.536 \pm 0.038)\,{\rm g\,cm}^{-3}$.

Since we use multipole expansion, our orbital model sensitively depends on the orientation of the central body.
Even a small change in the pole's coordinates can completely change the satellite's orbits.
In fact, we obtained more precise pole coordinates,
$\lambda_{\text{pole}} = (188.3 \pm 2.6)\,{\rm deg}$ and
$\beta_{\text{pole}} = (-88.2 \pm 0.2)\,{\rm deg}$,
from our orbital model than from the shape model alone.
Consequently, we derived an alternative shape model,
as discussed in Sect.~\ref{sec:shape}.

To demonstrate that these more precise pole coordinates are indeed preferable, we present a counterexample model in Table~\ref{tab:parameters}. This model is based on a rotational pole given by the ADAM algorithm in Sect.~\ref{sec:shape} and has a much higher $\chi^2$ value, making it unviable. 

\begin{table}[h]
\caption{
Parameters of the best-fit model, given with uncertainties. Also, for completeness, the parameters of a counter-example model based on the rotational pole given by the ADAM algorithm in Sect.~\ref{sec:shape}.
}
\label{tab:parameters}
\centering
\footnotesize
\begin{tabular}{*{4}{l}}
\hline\hline
& & best-fit model & counter-example \\
& & & model \\
\hline
$\chi_{\text{sky}}^2 + \chi_{\text{sky2}}^2$ & & $872$ & $19903$ \\
$N_{\text{measurements}}$ & & $837$ & $837$ \\
\hline \hline
variable & unit & value with uncertainty & value \\
\hline
$\lambda_{\text{pole}}$ & deg & $188.3 \pm 2.6$ & $68.5$ \\
$\beta_{\text{pole}}$ & deg & $-88.2 \pm 0.2$ & $-88.9$ \\
$M_{\text{total}}$ & $M_S$ & $(3.322 \substack{+0.003 \\ -0.006}) \times 10^{-12}$ & $3.323 \times 10^{-12}$ \\
$q_1$ & 1 & $(1.02 \pm 0.55) \times 10^{-6}$ & $0.97 \times 10^{-6}$ \\
$q_2$ & 1 & $(5.2 \pm 5.4) \times 10^{-7}$ & $2.7 \times 10^{-7}$ \\
$q_3$ & 1 & $(2.74 \pm 0.12) \times 10^{-5}$ & $5 \times 10^{-5}$ \\
\hline
$P_1$ & day & $1.2127 \pm 0.0002$ & $1.2127$ \\
$e_1$ & 1 & $0.028 \pm 0.005$ & $0.068$ \\
$i_1$ & deg & $179.7 \pm 0.6$ & $191.9$ \\
$\Omega_1$ & deg & $275.3 \pm 11.9$ & $265.2$ \\
$\varpi_1$ & deg & $324.3 \pm 18.5$ & $25.8$ \\
$\lambda_1$ & deg & $343.2 \pm 24.1$ & $347.9$ \\
\hline
$P_2$ & day & $1.6421 \pm 0.0004$ & $1.6420$ \\
$e_2$ & 1 & $0.064 \pm 0.007$ & $0.060$ \\
$i_2$ & deg & $183.7 \pm 0.3$ & $184.1$ \\
$\Omega_2$ & deg & $197.1 \pm 7.5$ & $194.7$ \\
$\varpi_2$ & deg & $107.9 \pm 12.1$ & $65.5$ \\
$\lambda_2$ & deg & $248.5 \pm 15.2$ & $244.9$ \\
\hline
$P_3$ & day & $5.30032 \pm 0.00015$ & $5.30041$ \\
$e_3$ & 1 & $0.123 \substack{+0.004 \\ -0.002}$ & $0.156$ \\
$i_3$ & deg & $175.3 \pm 0.2$ & $173.8$ \\
$\Omega_3$ & deg & $133.5 \pm 2.1$ & $129.7$ \\
$\varpi_3$ & deg & $356.8 \substack{+4.3 \\ -1.3}$ & $6.6$ \\
$\lambda_3$ & deg & $291.1 \substack{+4.3 \\ -0.1}$ & $289.1$ \\
\hline
\end{tabular}
\tablefoot{
The listed $\chi^2$ is with the weight $w_{\rm ao}$ set to zero and $N_{\text{measurements}}$ is the number of all astrometry measurements, both absolute and relative.
Orbital elements are referred to the epoch $T_0 = 2457021.567880$ (TDB).
Uncertainties are given at 1-$\sigma$.
$\lambda_{\text{pole}}$ and $\beta_{\text{pole}}$ are the ecliptic longitude and latitude of Elektra’s rotational pole.
$M_{\text{total}}$ is the total mass of the system,
$q_1 = m_1 / m_0$,
$q_2 = m_2 / (m_0 + m_1)$,
$q_3 = m_3 / (m_0 + m_1 + m_2)$ are the mass ratios,
where $m_0$ denotes the mass of the central body (i.e. Elektra),
$m_1$ body 1 (inner moon S/2014 1),
$m_2$ body 2 (middle moon S/2014 2),
$m_3$ body 3 (outer moon S/2003),
$P_1$ the orbital period of the inner orbit,
$e_1$ eccentricity,
$i_1$ inclination,
$\Omega_1$ longitude of the nodes,
$\varpi_1$ longitude of the pericentre,
$\lambda_1$ true longitude,
and the same for the middle and outer orbits.
}
\end{table}

\begin{table}[h]
\caption{
Dependent parameters of the best-fit model, given with uncertainties.
}
\label{tab:dependent_parameters}
\centering
\footnotesize
\begin{tabular}{*{3}{l}}
\hline \hline
variable & unit & value with uncertainty \\
\hline
$m_{0}$ & $M_S$ & $(3.322 \substack{+0.003 \\ -0.006}) \times 10^{-12}$ \\
$m_{1}$ & $M_S$ & $(3.4 \pm 1.8) \times 10^{-18}$ \\
$m_{2}$ & $M_S$ & $(1.7 \pm 1.8) \times 10^{-18}$ \\
$m_{3}$ & $M_S$ & $(9.1 \pm 0.4) \times 10^{-17}$ \\
\hline
$a_1$ & km & $496.8 \pm 0.4$ \\
$a_2$ & km & $608 \pm 0.5$ \\
$a_3$ & km & $1328 \pm 0.9$ \\
\hline
\end{tabular}
\tablefoot{
$m$ are the masses of the individual bodies and
$a$ are the semi-major axes.
The same notation as in Table \ref{tab:parameters} is used.
}
\end{table}

\subsection{MCMC analysis}

The Markov Chain Monte Carlo \citep[MCMC,][]{robert2011short, tierney1994markov} is a method for randomly sampling probability distributions. To sample the distribution functions of the parameters of our orbital model we used the Python package emcee \citep{foreman2013emcee}, which is an implementation of the MCMC. 

We let all parameters of the model be free, including 
the rotational pole of Elektra, 
the total mass of the system, 
mass ratios of the moons $q_1$, $q_2$, $q_3$ 
and 18 orbital parameters, six for each orbit of the three moons. 

To initialize the chains, we started with the parameters of the best-fit model and 
set some apriori limits to keep the walkers around the local minimum of our preferred solution. 
The limits were an order of magnitude higher and lower for $q_1$, $q_2$ \& $q_3$
to test if they are constrained and a range of a few hours for the orbital periods. 
For other variables, the limits are relevant only in the case of $\Omega_1$ 
and $\lambda_1$, because some part of their distribution lies beyond their upper limits, 
as can be seen in Fig.~\ref{fig:corner}. 
However, this should not notably affect the results. 

We ran the MCMC with 48 walkers (twice 24; the number of free parameters) for 4000 iterations. 
But, since our limits on orbital periods were not narrow enough, 
about 15 walkers ended up migrating into neighbouring local minima, 
so, without the loss of generality, we removed them. 
We set the burn-in phase as 2000 iterations,
after that, almost all parameters were in a steady state (Fig.~\ref{fig:chain}), 
except for $q_1$, $q_2$, $\log e_2$ and $\Omega_2$ (thus $\varpi_2$ \& $\lambda_2$), 
which still have some overall upward or downward trends. 
Regardless, this again should not notably affect the results.  

From the MCMC, we obtained the uncertainties of the parameters.
They are given in Table~\ref{tab:parameters} at 1-$\sigma$,
also known as the 16\% and 84\% percentiles.
In particular, we have a very precise determination of the total mass,
because it was constrained by all three orbits.  

The complete corner plot is plotted in Fig.~\ref{fig:corner}.
Some of the parameters are correlated with each other.
First, we have a strong positive triple correlation 
in each set of $\Omega_i$, $\varpi_i$ and $\lambda_i$, where $i = 1,2,3$.
This is not surprising given the definitions of these parameters.
The third triplet also has a positive correlation with $\log e_3$.
This could be explained by a specific observation geometry during the respective epochs.
The $\log e_3$ also has a negative correlation with $i_3$,
which could be due to the same reason.
Lastly, the first triplet has a positive correlation with the mass ratio $q_3$.     


\subsection{The third moon, S/2014 (130) 2}

The Kepler solution presented in \citep{Berdeu_2022A&A...658L...4B}
is a short-period (0.67\,d) orbit of the third moon S/2014~(130)~2.
However, it appears to `cross' the orbit of the inner moon S/2014~(130)~1.
Here, we present a 4-body model, which includes
mutual interactions,
multipoles ($\ell = 2$),
and external tides.
We surveyed all possible periods and
tested both inner and outer orbits.
In particular, we were interested in a coorbital solution
because the astrometric positions of S/2014~(130)~2 are always close to the orbit of S/2014~(130)~1.

\begin{figure}
\centering
\includegraphics[width=8.7cm]{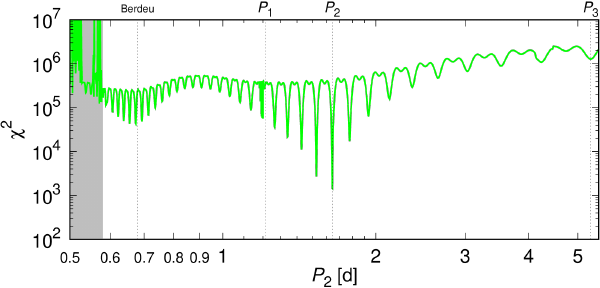}
\caption{
Periodogram for the orbit of the third moon (S/2014~(130)~2).
The $\chi^2$ values were computed from the 2014 astrometric data only,
without contributions for other moons.
The osculating period $P_2$ was varied,
other elements were kept fixed.
At the epoch $T_0 = 2457021.567880$ (TDB),
the synthetic position was very close to the observed one.
The best-fit period is $1.642\,{\rm d}$.
For comparison, the period of $0.678\,{\rm d}$ 
from \citet{Berdeu_2022A&A...658L...4B} 
is indicated.
}
\label{130_test8_2014__254_map}
\end{figure}

The periodogram for the third moon is presented in Fig.~\ref{130_test8_2014__254_map}.
Its orbit is unstable below 0.58\,d;
0.67\,d is only one local minimum, but undeniably not the deepest.
The coorbital solutions have periods between 1.19 and 1.20\,d,
but none of them is deep enough, unfortunately.
A regular series of minima is seen for longer periods,
with 1.64\,d being the deepest minimum.

This minimum is our preferred solution for S/2014~(130)~2.
It is a stable orbit,
close to the mean-motion resonance with S/2014~(130)~1.
The ratio of periods 1.35 is not exactly 4/3,
because of the respective critical angle:
\begin{equation}
\sigma = 4\lambda_2 - 3\lambda_1 + \varpi_1
\end{equation}
or alternatively:
\begin{equation}
\sigma' = 4\lambda_2 - 3\lambda_1 + \varpi_2
\end{equation}
includes a non-negligible precession contribution.
regardless, neither of these angles librates;
a necessary period for an exact resonance is slightly shorter,
$P_2 = 1.608\,{\rm d}$.
Nonetheless, the proximity to the mean-motion resonance 
might be an independent indication of a correct solution
(such as for (216)~Kleopatra; \citealt{brovz2021advanced}).

We also verified the influence of moon masses on their orbital evolution.
According to our tests with mass ratios up to $10^{-4}$,
mutual perturbations are weak,
compared to the remaining systematics in astrometry.
Therefore, the moon masses remain unconstrained
and the mass ratios inferred from photometry
(approximately $10^{-6}$, $5\times10^{-7}$, $3\times10^{-5}$)
should be preferred.

\begin{figure}
\centering
\includegraphics[width=9cm]{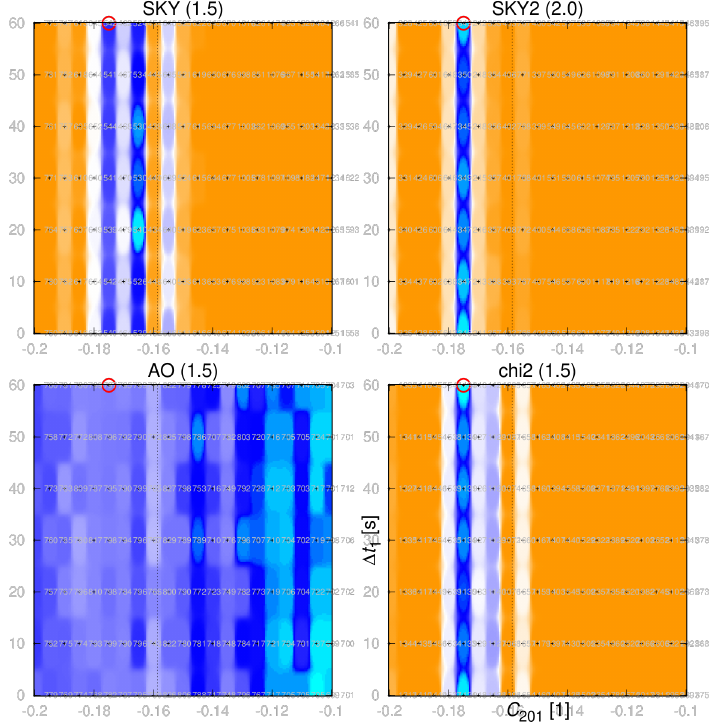}
\caption{
Overview of best-fit models of the (130) system
for the pair of fixed parameters:
the oblateness $C_{20,1}$ and
the tidal time lag $\Delta t_1$ \citep{brovz2022observed} of the central body.
All other parameters were free.
Each of the 147 models
was converged for up to 1000 iterations,
that is 147000 models in total.
The $\chi^2$ values are plotted as colours:
\color{cyan}cyan\color{black}\ corresponds to the overall best-fit,
\color{blue}blue\color{black}\ to good fits (1.2 times the best fit),
\color{orange}orange\color{black}\ to poor fits (1.5).
The four panels correspond to the $\chi^2$ contributions:
astrometry (SKY),
relative astrometry of the moons (SKY2),
silhouettes of the central body (AO),
and the weighted sum of them.
}
\label{130_fitting4_6thorder__920_C201_Deltat1_ALL}
\end{figure}


\subsection{Dynamical constraints for the oblateness}\label{sec:C20}

Having all three orbits,
we performed a fitting of the oblateness $C_{20}$ of the central body.
We were interested not only in the best-fit model
but also in bad fits,
which would enable us to reject the respective models.
We computed an extended grid of models,
with the oblateness $C_{20}$ and the time lag $\Delta t$ parameters kept fixed,
while all other parameters were set free.

We included all multipole terms up to the order $\ell = 6$
(from Table~\ref{tab:multipole_coef})
to assert that none of these high-order contributions to the total precession rate is missed.
We also verified that orders $\ell > 6$ only negligibly affect the value of $\chi^2$.
Initial conditions for the simplex algorithm \citep{nelder1965simplex} must be set up precisely,
especially $P_1$, $P_2$, $P_3$ must be close to the true/global minimum
(for a given value of $C_{20}$),
otherwise, the simplex could be `stuck'.
We used two fully-converged models for two different values of $C_{20}$,
we verified these minima are global by a $\chi^2$ mapping,
knowing the typical spacing between the local minima, such as
$\Delta P = P_1^2/(22.1\,{\rm d}) \doteq 0.066\,{\rm d}$, or
$\Delta P' = P_1^2/(1701\,{\rm d}) \doteq 0.00086\,{\rm d}$,
depending on the time span (2014 only, 2014--2019).
Eventually, we determined the linear relations
$P_1(C_{20})$, $P_2(C_{20})$, $P_3(C_{20})$,
and interpolated the periods accordingly.

Overall, the best-fit model has $\chi^2 = 932$,
with contributions
$\chi^2_{\rm sky} = 479$,
$\chi^2_{\rm sky2} = 211$,
$\chi^2_{\rm ao} = 792$,
where we used the weight $w_{\rm ao} = 0.3$.
All models are summarised in
Fig.~\ref{130_fitting4_6thorder__920_C201_Deltat1_ALL}.
The best-fit value of $C_{20} \simeq -0.18$,
with an uncertainty of less than $0.01$.
This is slightly larger than the value computed for a homogeneous body.

Most importantly, we did not find any fits with $C_{20} \simeq -0.16$,
which have lower or comparable $\chi^2$.
On the contrary, the values were always significantly higher.
This might be an indication of irregular,
or (partially) differentiated internal structure,
however, the irregularity should be more oblate, not more spherical.

\paragraph{Possible Elektra family.}
Given the meteorite analogue material (CM chondrites)
and its average density ($2.13\,{\rm g}\,{\rm cm}^{-3}$),
it implies a substantial porosity of $28\,\%$ for Elektra.
If true, a reaccumulation event might be at the origin
of Elektra itself (and its satellite system).
Usually, the event ends with a reaccumulation of streams
(see, e.g. \citealt{Broz_2022A&A...664A..69B}),
which deposit loose, low-dense, rubble-pile material
--- creating `hills', some of which could be observed at the limb
(see, e.g., Fig.~\ref{fig:comparison}, top row, 5th column).
If true, fragments from this event should also form a family.
Even though there is no `official' Elektra family
\citep{Nesvorny_2015aste.book..297N},
(130) Elektra is embedded in the Alauda family ((702), FIN 902 \footnote{Family Identification Number = 902 \citep{Nesvorny_2015aste.book..297N}}).
Given all the arguments above, a part of this family should belong to Elektra,
in particular, bodies with lower semimajor axis ($a < 3.07\,{\rm au}$),
which do have similar inclinations as (130) Elektra ($\sin i \simeq 0.38$).


\section{Conclusions}

We presented the first self-consistent model for all three moons of (130) Elektra.
The model covers a considerably long time span (2014--2019) requiring a sufficiently complex dynamical model, including at least oblateness ($\ell = 2$), which induces precession. 
With the constraint of three orbits and assuming a homogeneous internal structure, we obtained the precise mass of Elektra $(6.606 \substack{+0.007 \\ -0.013}) \times 10^{18}\,{\rm kg}$.

The relative astrometry of the moons seems more reliable than measurements with respect to (130) Elektra since any possible issues with determining the photocentre of the primary are absent.
The relative astrometry strongly suggests a dynamical oblateness of $J_2 \simeq 0.18$, even with the high-order ($\ell = 6$) multipoles included, which also contribute to the total precession.
This higher oblateness would require, either some internal structure
or a substantial modification of the shape, ‘enforced' by the observed precession.
In the future, this should lead to new `photo-dynamical' shape reconstruction methods.


\begin{acknowledgements}
The Czech Science Foundation supported this work through grants GA22-17783S (M. Fuksa, J.H.) and GA21-11058S (M.B.).
In this work, measurements from the BlueEye600 telescope, supported by Charles University, were used.
\end{acknowledgements}





\clearpage
\newpage

\begin{appendix}

\section{MCMC figures}

\noindent\begin{minipage}{\textwidth}
\centering
   \includegraphics[width=18.5cm]{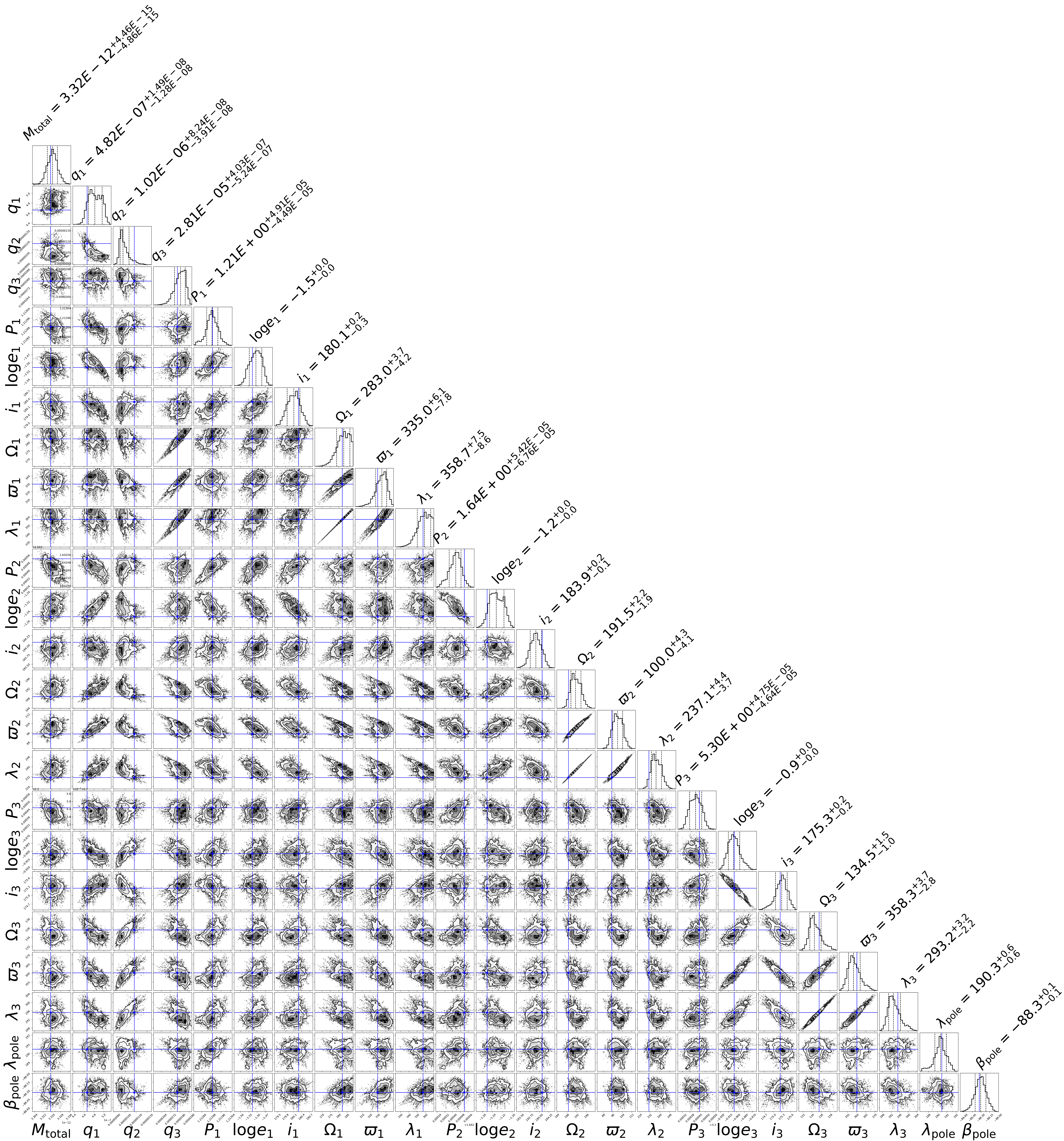}
     \captionof{figure}{Corner plot showing the probability distributions of all parameters of the orbital model. \emph{Blue} lines indicate a best-fit parameter, while the three \emph{dashed} lines show the 16\%, 50\% and 84\% percentiles for each parameter.}
     \label{fig:corner}
\end{minipage}

\begin{figure*}
\centering
   \includegraphics[width=18.0cm]{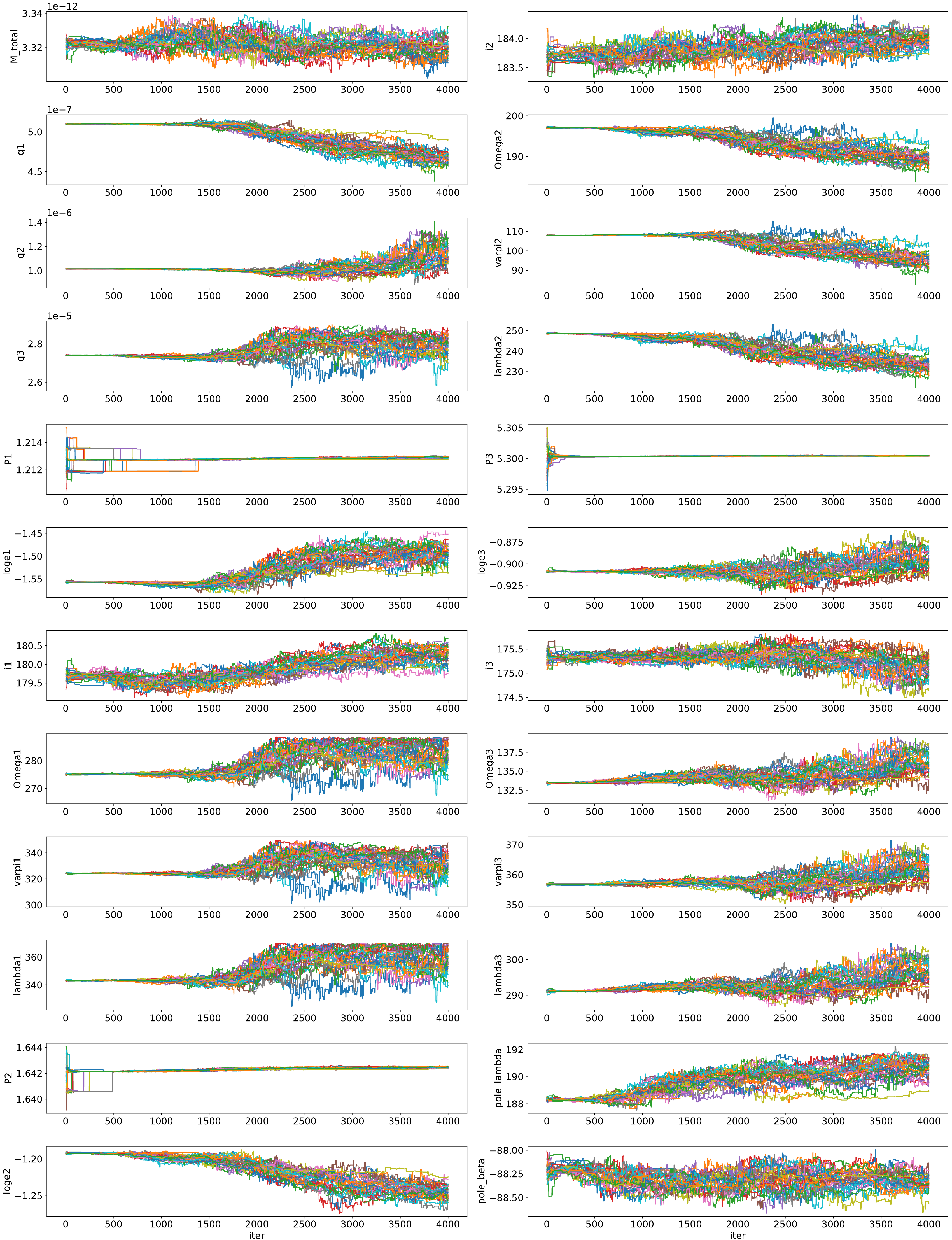}
     \caption{A plot of the Markov chains. Each subplot corresponds to one variable and contains 33 walkers, each denoted by a different colour. Walkers, over 4000 iterations, mapped out possible values of the variables.}
     \label{fig:chain}
\end{figure*}

\FloatBarrier

\section{Observation data}
\label{app:observation_data}

In this appendix, we present three tables of astrophotometry fits of the three moons of Elektra. In each table, $x$ and $y$ are the Cartesian positions of the moons relative to the photocentre of Elektra.

\raggedbottom

\begin{table}[H]
\caption{Astrophotometry fit of the inner moon S/2014 (130) 1.}
\label{table:astro_fit_1}     
\centering                        
\begin{tabular}{c c c}         
\hline\hline           
July 2019 (Julian date) & $x$ (mas) & $y$ (mas) \\  
\hline                        
2458694.861418 & 172.9 & 36.8 \\
2458694.864054 & 177 & 37.5 \\
2458694.866710 & 180 & 43.6 \\
2458694.875133 & 188.7 & 46.5 \\
2458694.877797 & 192.8 & 46.4 \\
2458694.880464 & 196.2 & 47.4 \\
2458694.883109 & 200.8 & 48.2 \\
2458694.885767 & 207 & 49.7 \\
\hline             
August 2019 (Julian date) & $x$ (mas) & $y$ (mas) \\  
\hline 
2458699.760132 & 344.8 & 96.2 \\
2458699.762773 & 346.2 & 99.7 \\
2458699.765448 & 345.7 & 101.1 \\
2458699.768099 & 349.2 & 105.4 \\
2458699.770757 & 345.2 & 103.2 \\
2458700.887314 & 275 & 83.1 \\
2458700.897933 & 295.2 & 87 \\
2458701.728692 & -351.6 & -120.7 \\
2458701.731355 & -341.9 & -114.9 \\
2458701.734024 & -339.6 & -115.9 \\
2458701.736668 & -348.2 & -121.6 \\
2458701.739317 & -337.1 & -118 \\
\hline                                       
\end{tabular}
\end{table}

\begin{table}[H]
\caption{Astrophotometry fit of the middle moon S/2014 (130) 2.}
\label{table:astro_fit_2}     
\centering                        
\begin{tabular}{c c c}         
\hline\hline           
July 2019 (Julian date) & $x$ (mas) & $y$ (mas) \\  
\hline                        
2458694.856099 & 366.6 & 169.1 \\
2458694.858750 & 371.2 & 161.5 \\
2458694.861418 & 377.6 & 160.5 \\
2458694.864054 & 370.1 & 153.6 \\
2458694.866710 & 362.1 & 153.6 \\
2458694.875133 & 352.9 & 148.3 \\
2458694.877797 & 353.1 & 148.8 \\
2458694.880464 & 348.5 & 145.1 \\
2458694.883109 & 349.4 & 153.9 \\
2458694.885767 & 342.4 & 146.6 \\
\hline          
August 2019 (Julian date) & $x$ (mas) & $y$ (mas) \\  
\hline 
2458699.760132 & 340.2 & 154.7 \\
2458699.762773 & 341.4 & 149.8 \\
2458699.765448 & 340.2 & 153.2 \\
2458699.768099 & 331.5 & 152.3 \\
2458699.770757 & 321.3 & 148.2 \\
2458701.728692 & -228 & -79.3 \\
2458701.731355 & -231.2 & -83.3 \\
2458701.734024 & -233 & -90.9 \\
2458701.736668 & -246.6 & -88.6 \\
2458701.739317 & -247.7 & -91.7 \\
\hline                                       
\end{tabular}
\end{table}

\begin{table}[H]
\caption{Astrophotometry fit of the outer moon S/2003 (130) 1.}
\label{table:astro_fit_3}     
\centering                        
\begin{tabular}{c c c}         
\hline\hline           
July 2019 (Julian date) & $x$ (mas) & $y$ (mas) \\  
\hline                        
2458694.85737 & -663.6 & -233.7 \\
2458694.86799 & -654 & -233.9 \\
2458694.87641 & -651 & -230 \\
2458694.88704 & -645.3 & -227 \\
\hline           
August 2019 (Julian date) & $x$ (mas) & $y$ (mas) \\  
\hline
2458699.69701 & -915.2 & -385.7 \\
2458699.70762 & -918.9 & -388.3 \\
2458699.76141 & -898.5 & -382.2 \\
2458699.77203 & -898.1 & -379.5 \\
2458700.88859 & 162.5 & 137.4 \\
2458700.89921 & 188.4 & 147.5 \\
2458701.72997 & 956.5 & 443.4 \\
2458701.74059 & 960.9 & 445.7 \\
\hline                                       
\end{tabular}
\end{table}

\section{Extension of the best-fit orbital model}

\begin{figure}[H]
  \resizebox{\hsize}{!}{
    \includegraphics[width=\textwidth]{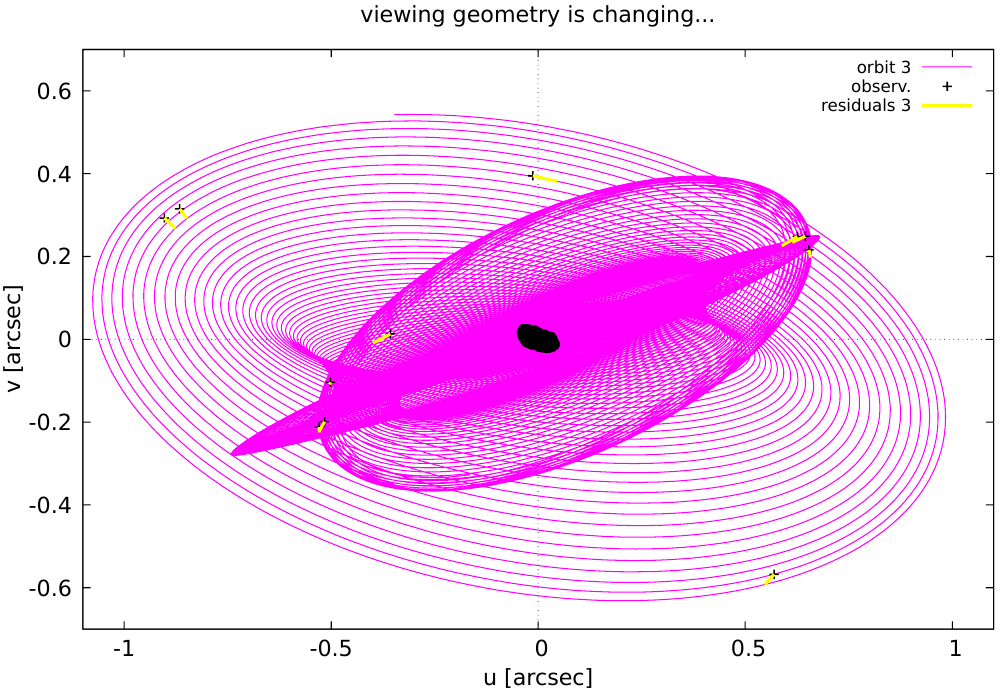}
    }
  \caption{
     Extension of the best-fit quadrupole model, now with $\chi^2$ = 1406. Shown is only the relevant time-span of 903 days for orbit 3.
      The orbit is plotted in the (\emph{u; v}) coordinates (\emph{pink} line),
      together with observed positions (\emph{black crosses}),
     and residuals (\emph{yellow} lines).
     Elektra’s shape for one of the epochs is overplotted in \emph{black}.
  }
  \label{fig:graph_of_model_Marchis}
\end{figure}

In this section, we present an extension of the best-fit orbital model from Sect.~\ref{sec:orbit}.
There are several astrometric measurements of the outer moon S/2003,
from 2003 to 2006, that were published in \citet{marchis2008main}.
To further test our best-fit model we extended it by about 10 years into the past
and added these older measurements.
With some additional converging the resulting model has a $\chi^2$ = 1406.
In Fig.~\ref{fig:graph_of_model_Marchis}, we present the fit of the outer moon's orbit
to the older positions.

\newpage

\section{Lightcurves}
\label{app:lightcurves}

\noindent\begin{minipage}{\textwidth}
\centering
\includegraphics[width=15cm]{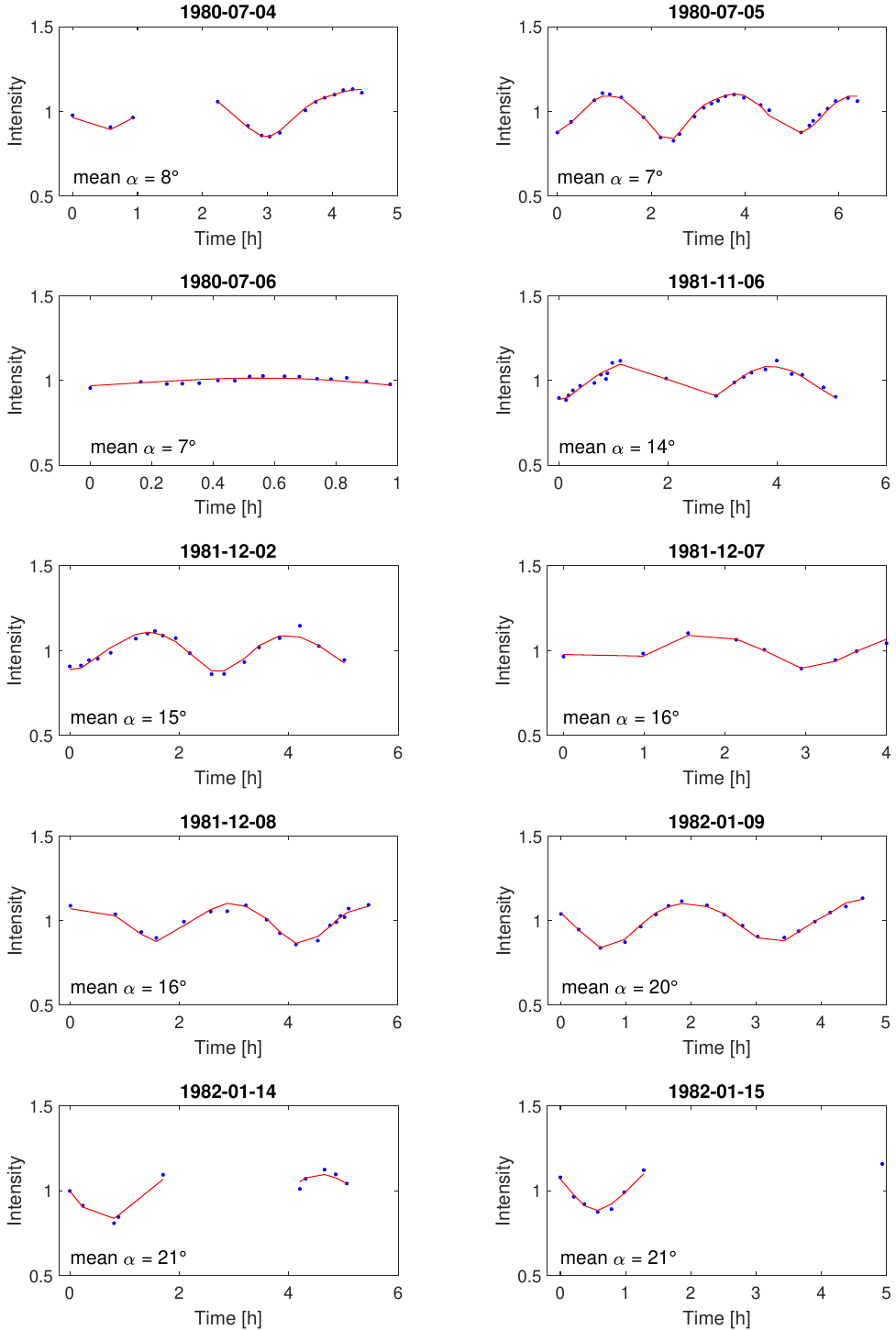}
\captionof{figure}{The lightcurves used in the shape reconstruction compared to the simulated lightcurves of the best-fit model. The observed intensity is represented by \emph{blue dots}, while the simulated intensity of the shape model is depicted in \emph{red}. The mean phase angle $\alpha$ for each observation is provided in the bottom left corner of each graph.
}
\label{fig:Lightcurves_1}
\end{minipage}

\addtocounter{figure}{-1}

\begin{figure*}
\centering
\includegraphics[width=16cm]{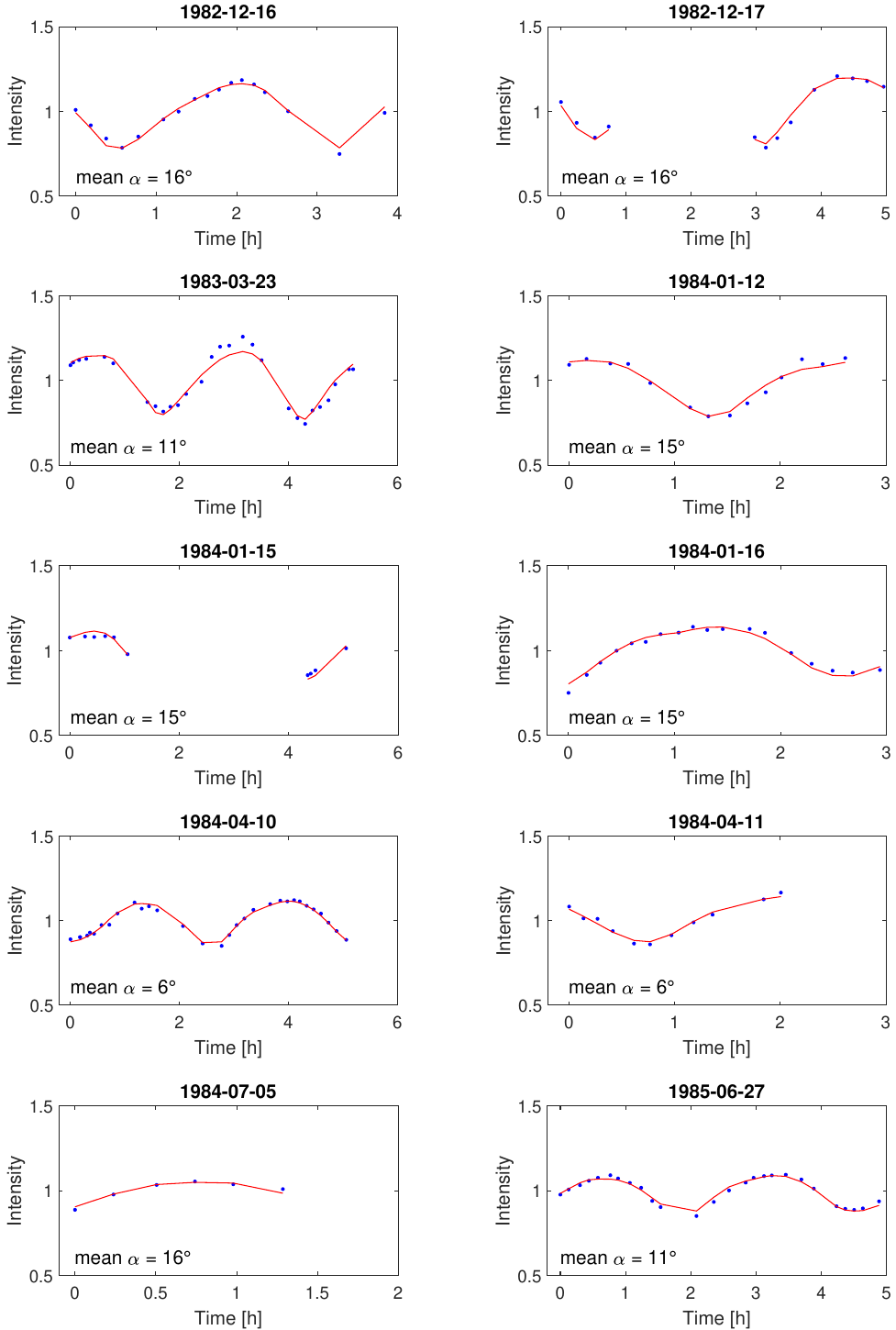}
\captionof{figure}{Continued.}
\label{fig:Lightcurves_2}
\end{figure*}

\addtocounter{figure}{-1}

\begin{figure*}
\centering
\includegraphics[width=16cm]{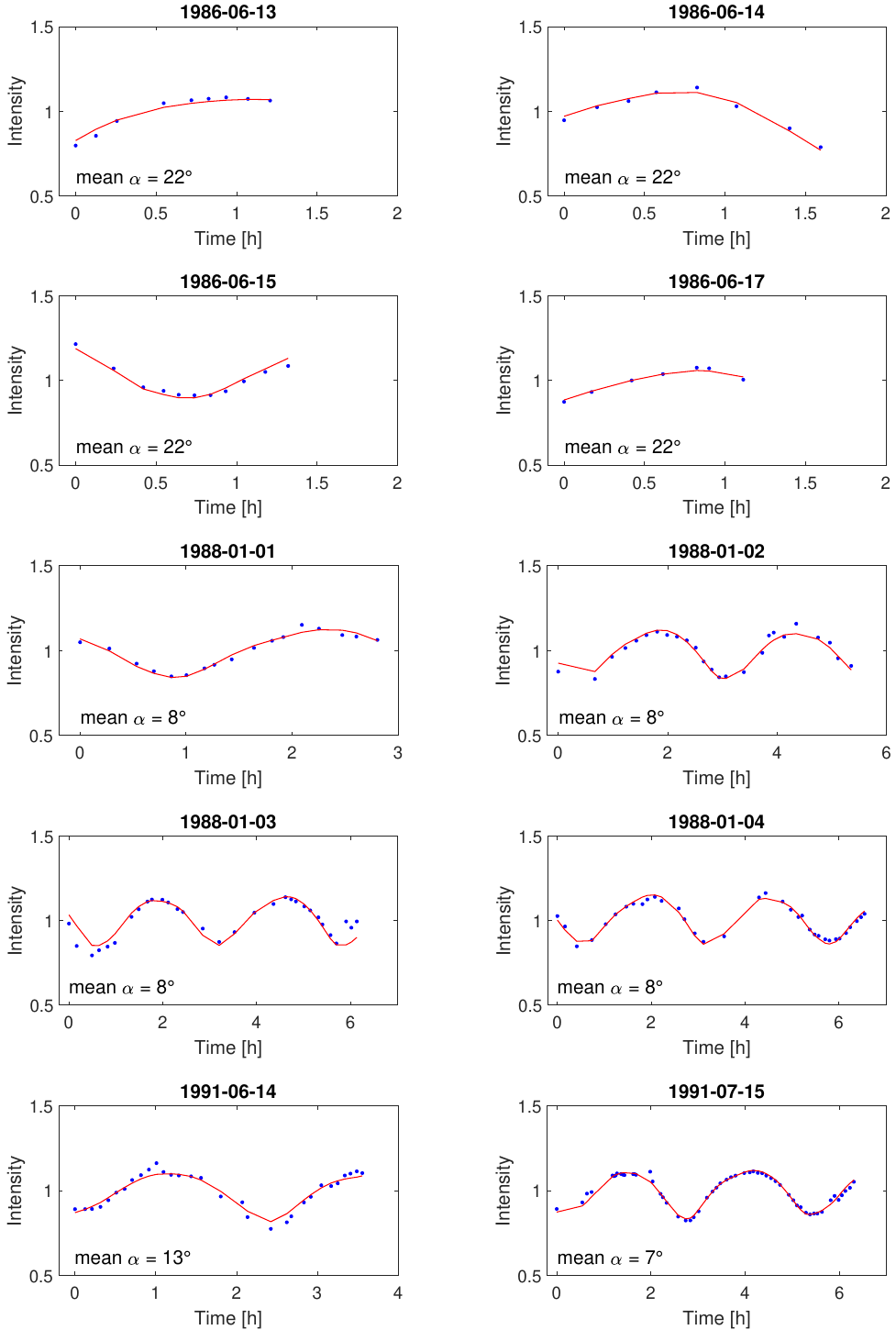}
\captionof{figure}{Continued.}
\label{fig:Lightcurves_3}
\end{figure*}

\addtocounter{figure}{-1}

\begin{figure*}
\centering
\includegraphics[width=16cm]{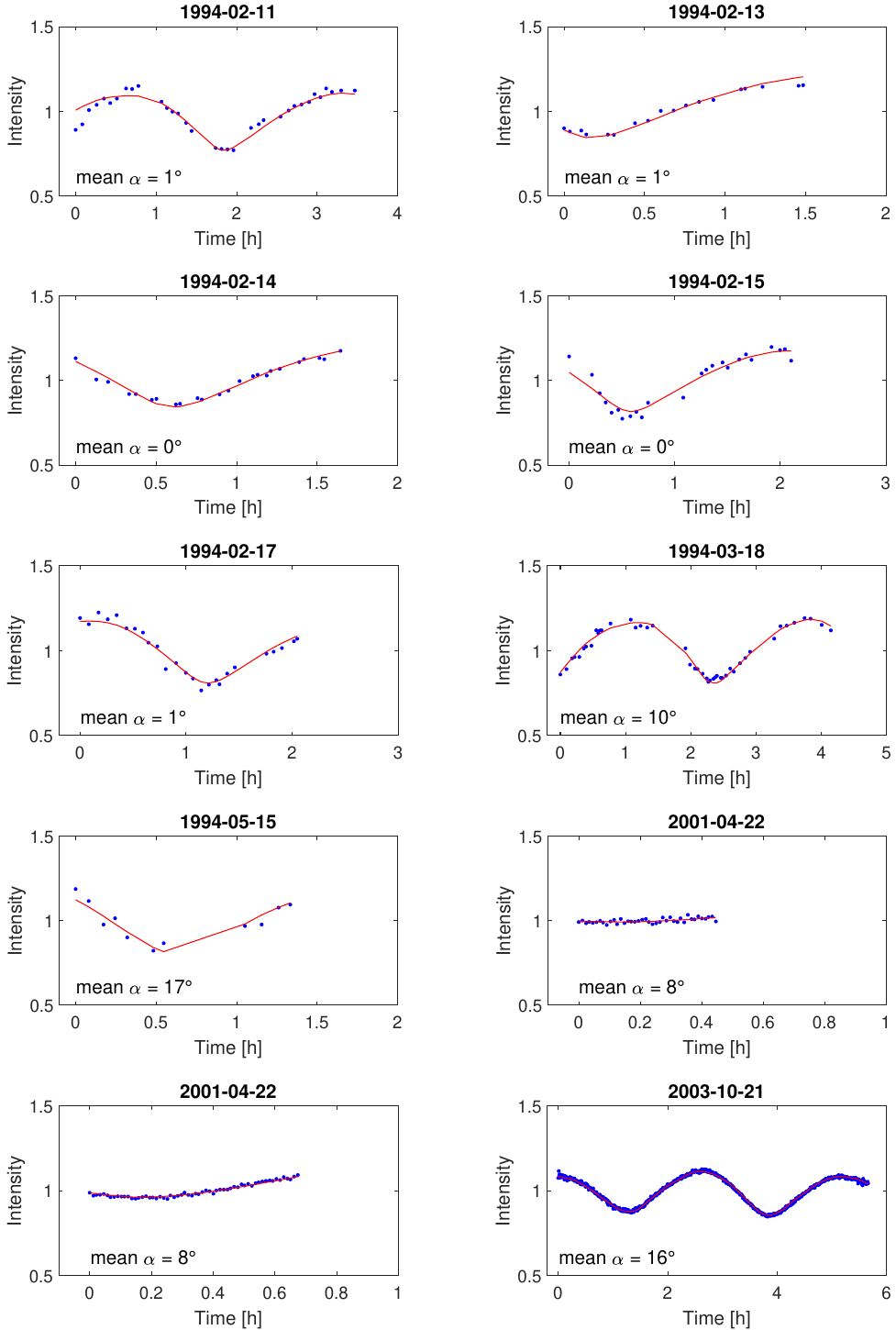}
\captionof{figure}{Continued.}
\label{fig:Lightcurves_4}
\end{figure*}

\addtocounter{figure}{-1}

\begin{figure*}
\centering
\includegraphics[width=16cm]{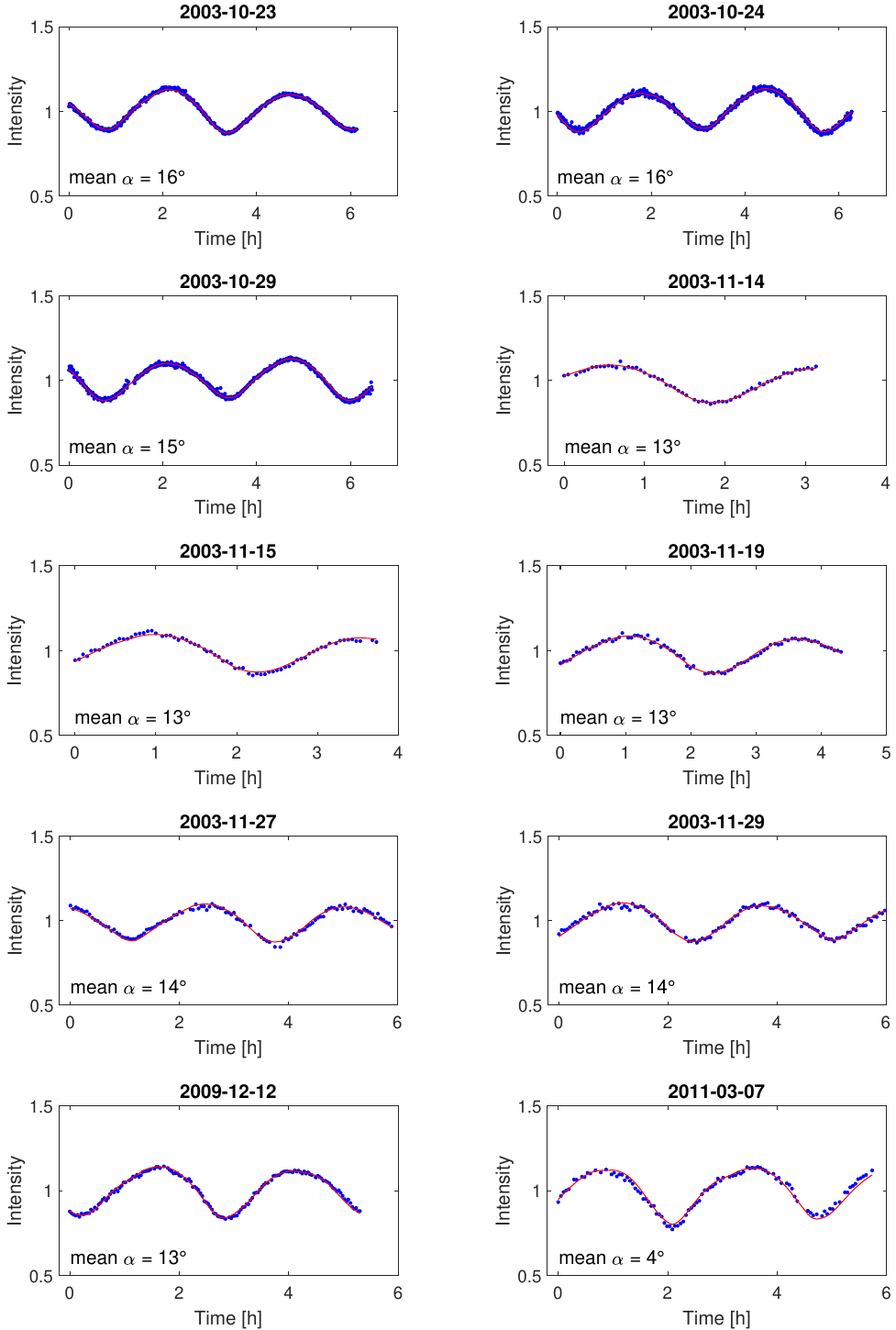}
\captionof{figure}{Continued.}
\label{fig:Lightcurves_5}
\end{figure*}

\addtocounter{figure}{-1}

\begin{figure*}
\centering
\includegraphics[width=16cm]{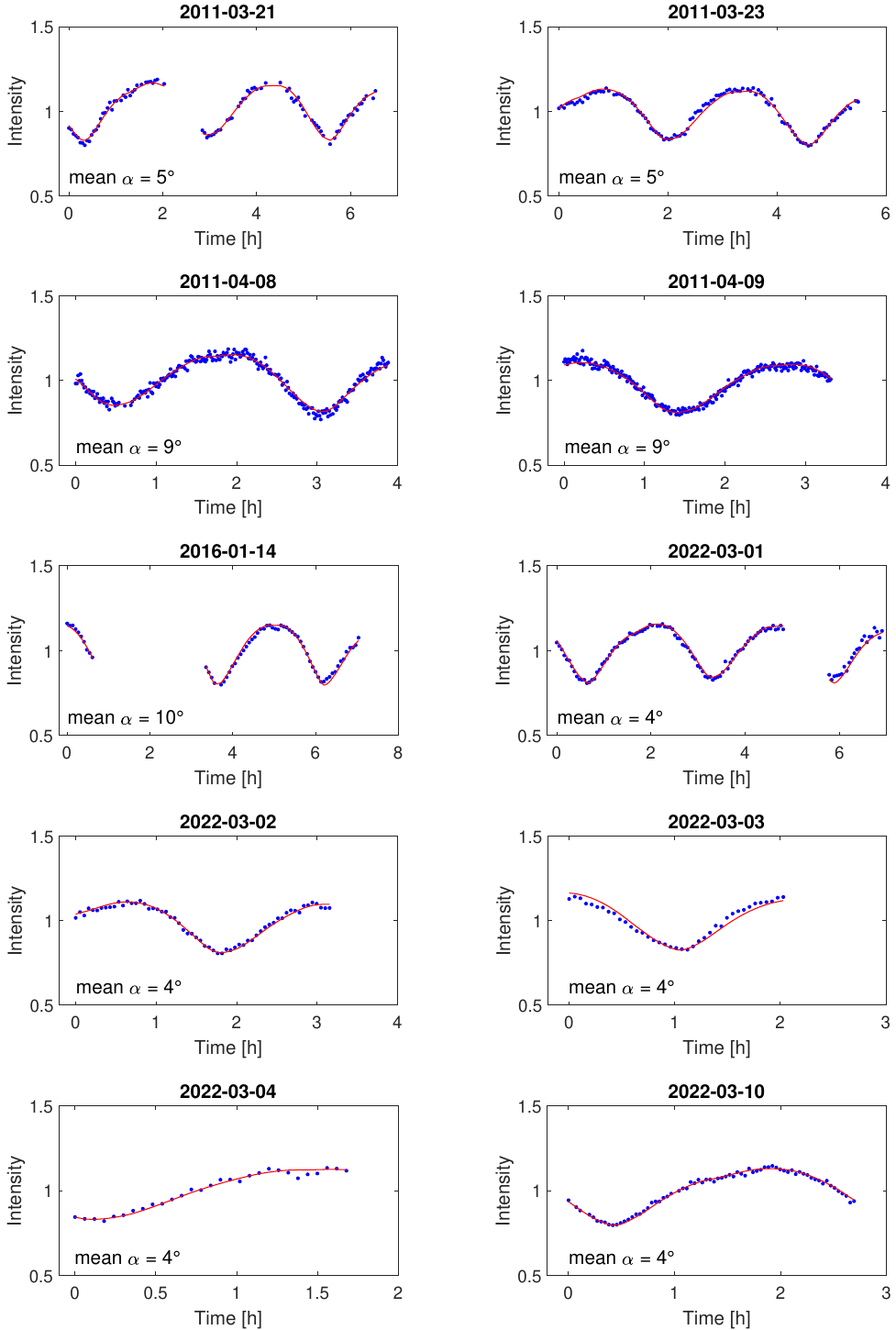}
\captionof{figure}{Continued.}
\label{fig:Lightcurves_6}
\end{figure*}

\end{appendix}
\end{document}